\documentclass[11pt,a4paper]{article}
\pdfoutput=1
\usepackage[utf8]{inputenc}
\usepackage[portuguese, english]{babel}
\usepackage{jheppub}
\usepackage{hyperref}
\usepackage{amsmath}
\usepackage{amsfonts}
\usepackage{amssymb}
\usepackage{graphicx}
\usepackage{caption}
\usepackage{subcaption}
\title{Universal Aspects of $U(1)$ Gauge Field Localization on Branes in $D$-dimensions}
\author[a]{L. F. F. Freitas,}
\author[a]{G. Alencar} %
\author[a]{and R. R. Landim}
\affiliation[a]{Universidade Federal do Ceará - UFC,\\
Departamento de Física, Universidade Federal do Ceará, 60451-970 Fortaleza, Ceará, Brazil}
\emailAdd{luizfreitas@fisica.ufc.br}
\emailAdd{geovamaciel@gmail.com}
\abstract{\boldmath  In this work, we study the general properties of the $D$-vector field localization on $(D-d-1)$-brane with co-dimension $d$. We consider a conformally flat metric with the warp factor depending only on the transverse extra dimensions. We employ the geometrical coupling mechanism and find  an analytical solution for the $U(1)$ gauge field  valid for any warp factor. Using this solution we find that the only condition necessary for  localization is that the bulk geometry is asymptotically AdS. Therefore, our solution has an universal validity for any warp factor and is independent of the particular model considered. We also show that the model has no tachyonic modes. Finally, we study the scalar components of the $D$-vector field. As a general result, we show that if we consider the coupling with the tensor and the Ricci scalar in higher co-dimensions, there is an indication that both sectors will be localized. As a concrete example, the above techniques are applied for the intersecting brane model. We obtain that the branes introduce boundary conditions that fix all parameters of the model in such a way that both sectors, gauge and scalar fields, are confined.}

\arxivnumber{1809.07197}

\begin{document}
\maketitle
\section{Introduction}\label{Sec-0}
The formulation of models in spacetimes with more than $4$-dimensions as a tool to solve problems in physics is not new \cite{Nordstrom, Kaluza}. However, only after the development of string theory and the compactification mechanisms of extra dimensions, at the end of the last century, this tool began to be regarded a possible real description of nature \cite{Akama}. A feature of these higher dimensional theories was the need of the extra dimensions to be compactified into a very small spatial volume inaccessible in the available energy range. This because the Newton's gravitational law depends explicitly on the number of spatial dimensions, and it indicates the presence of only three large spatial dimensions. The first to speculate about the possibility of these extra dimensions being non-compact were Rubakov and Shaposhnikov \cite{Rubakov}. The authors showed that a large extra dimension is allowed as long as the fields of the Standard Model (SM), as well as the gravity, are confined to a spatial $3$-dimensional hypersurface ($3$-brane). In such a way that our energy scale (TeV)  does not allow us to access them.

In this direction, L. Randall and R. Sundrum (RS) proposed two models with warped geometry in an $AdS_{5}$ spacetime with delta-like $3$-branes \cite{RS1, RS2}. The RS-I model, proposed to solve the Higgs hierarchy problem, considers a $5$-dimensional universe $(x^{\mu},\phi)$ with the spatial dimension $\phi$ compactified under a circle with an orbifold symmetry $S_{1}/Z_{2}$. At the fixed points $(\phi = 0, \pi)$ are placed two delta-like $3$-branes, and the $3$-brane at $\phi=\pi$ would correspond to our universe with all SM fields confined. The RS-II model considers only one delta-like $3$-brane with a non-compact and infinitely large extra dimension $(x^{\mu}, y)$, and it was proposed as an alternative to the compactification. On both models, the gravity and the scalar fields are localized on the $3$-brane. Despite this, the other fields of SM are not confined as expected by RS \cite{Chang, Oda1, Barut}. Soon after the success of RS models, other proposals of braneworld with localized gravity were presented in five dimensions ($5$D): smooth versions of RS-II (thick branes) \cite{Gremm1, Kehagias}; thick branes with inner structure \cite{Bazeia0, Bazeia1}; cosmological models, where the $3$-brane has a \textit{Robertson-Walker} metric \cite{Liu0, Guo}. Also, other solutions in spacetimes with more than $5$D as $3$-brane generated by string-like or vortex topological defects  in $6$D \cite{Gogberashvili1, Silva}; braneworld models generated by the intersection of delta-like branes \cite{Arkani}, and other proposals \cite{Cohen, Kim, Corradini, Cendejas, LiuFu}.

In all the above models, the issue of the SM fields localization is always an important point to be verified \cite{Daemi, Koley, Ringeval, Landim, Melfo, Chun}, mainly the confinement of $U(1)$ gauge field. It is a well known fact that the free abelian gauge field is not confined in $5$D braneworld models \cite{Pomarol, Davoudiasl}. In $6$D or higher dimensional models, the confinement seems to be possible since there is only one infinitely large extra dimension \cite{Midodashvili, Giovannini0}, however, a more detailed analysis, by exploring the Hodge duality symmetry \cite{Duff} for example, shows the opposite. Some attempts to solve this problem were performed. In most cases, introducing new degrees of freedom such as interaction terms with fermionic or scalar (\textit{dilatonic}) fields \cite{Oda2, Dvali, Betell, Guerrero, Araujo, Cruz0, Cruz1}. Although these mechanisms allow to confine the $U(1)$ gauge field, other questions arise: what is the meaning of these new fields for the theory? or, do such mechanisms work for other braneworld models?

Recently, Ghoroku and Nakamura (GN) developed a localization mechanism in RS-II model without the need of introducing new degrees of freedom \cite{Ghoroku}. They introduced a mass term and a non-covariant interaction between the vector field and the $3$-brane. This mechanism works, however, there is no solid motivation for the introduction of a coupling with the $3$-brane. Furthermore, they introduce a free parameter in the theory. Based on this method, a purely geometric localization mechanism was proposed in Refs. \cite{Flachi, Alencar, Zhen-Hua}, where an interaction term of the vector field with the Ricci scalar is added. This \textit{geometric coupling} allows us to localize the massless mode of the abelian vector field and has the advantage that it is covariant and does not introduce new degrees of freedom neither free parameters in the theory. Beyond that, the interaction with the $3$-brane in GN model arises as a consequence of the coupling of the vector field with the gravity in the bulk. This mechanism showed a very interesting and powerful feature. The massless mode solution has a shape that allows to confine a gauge field not only for the RS-II model, but also for other models where the brane is not delta-like.  Afterward, this mechanism was applied to the massive modes which the resonant modes of vector and $p$-form fields can be studied \cite{Alencar1, Alencar4, Alencar6}; also, by looking for phenomenological consequences, as a residual non-zero mass for the photon due to the existence of extra dimensions \cite{Alencar2}; beyond this, the application of this non-minimal coupling with gravity to analyze the localization of other fields \cite{Alencar3, Alencar5}. All this points were developed in warped models with only \textit{one} infinitely large extra dimension.

Despite the above results, a generalization of the geometrical coupling mechanism to more than one transverse extra dimensions is lacking. As presented above many other scenarios of braneworld with more extra dimensions were proposed allowing a more rich gravitational configuration. Furthermore, the vector field will have more scalar components which can play an important role on the brane. In view of all this, in this manuscript we study the vector field localization in a generic spacetime with an arbitrary number of large extra dimensions. First, we look for the universal aspects of localization for the two sectors of the $D$-vector field, the transverse $(D-d)$-vector field and the scalar components on the brane. By universal it means aspects that do not depends on any specific braneworld model, but only on the fact that they are asymptotically AdS. Also, due to the localization of such fields be valid for a wide variety of braneworld models. In this way, we look for the possibility of both components can be simultaneously localized for some range of the parameters of the model. We want to make it clear that the localization of the scalar components does not necessarily imply corrections to the Coulomb law, instead such components could be interpreted as Higgs fields, for example. As a concrete application of our results, we consider the intersecting brane model cited above.

This work is organized as follows: In section (\ref{Sec-3}), the general aspects of the confinement of the $D$-vector field on a generic braneworld scenario are discussed. We analyze in sections (\ref{Sec-3-1}) and (\ref{Sec-3-3}) the cases of the transverse $(D-d)$-vector field and the scalar components respectively. In section (\ref{Sec-5}), we carry out an application of our general results for the specific case of intersecting branes model. The conclusions are left for section (\ref{Sec-6}).

\section{\boldmath  Geometric Coupling as Universal Localization Mechanism for Vector Field}\label{Sec-3}
In this section, we will use the geometric couplings of the vector field with the scalar and the Ricci tensor in a generic braneworld model with arbitrary co-dimensions. In doing this, it will be shown that the geometric coupling has an universal validity as localization mechanism to the vector field. In a general way, when we talk about localization of fields in braneworld models, it means that we want to factor out the action
\begin{equation}
S=\int d^{4}x d^{D-4}z\sqrt{-g^{(D)}}\mathcal{L}_{\mbox{(matter)}}^{(D)},\label{Sec-1-02}
\end{equation}
into a sector containing an effective action on the $3$-brane and an integral in the coordinates of extra dimensions, {\it i.e.},
\begin{equation}
S=\int d^{D-4}z f(z)\int d^{4}x \sqrt{-g^{(4)}}\mathcal{L}_{\mbox{(matter)}}^{(4)}=K\int d^{4}x \sqrt{-g^{(4)}}\mathcal{L}_{\mbox{(matter)}}^{(4)}.\label{Sec-1-03}
\end{equation}
Thus, we say that the theory is well-defined, {\it i.e.}, the field is localized on the brane, when the integral $K$ is finite. As mentioned above, we will restrict ourselves to the localization of the vector field in a generic $D$-dimensional RS-like braneworld scenario. 

 Let us start by proposing the action for the $D$-vector field as 
\begin{eqnarray}
\! S_{2}\!=-\!\!\int \! d^{(D-d)}x d^{d}y \sqrt{-g}\left[\frac{1}{4}\mathcal{F}_{MN}\mathcal{F}^{MN}\!+\!\frac{\lambda_{1}}{2}R\mathcal{A}_{M}\mathcal{A}^{M}\!+\!\frac{\lambda_{2}}{2}R_{MN}\mathcal{A}^{M}\mathcal{A}^{N}\right]\!.\label{Sec-3-02}
\end{eqnarray}
In the above equation $\mathcal{A}_{M}(x,y)$ is the vector field in $D$-dimensions, $\mathcal{F}_{MP}=\partial_{M}\mathcal{A}_{P}-\partial_{P}\mathcal{A}_{M}$ is the field strength tensor and $R$ and  $R_{NM}$ are the scalar and Ricci tensor respectively. We should point that the gravitational field will be considered as a background and we are not interested in the backreaction or fluctuations of the geometry. This is justified since the vector field  is a small perturbation. As can be seen in Eq. (\ref{Sec-3-02}) above, the interaction between the vector field and geometry is cubic and that would contribute only to higher order corrections. Therefore the localization of the vector field can be carried out separately and we can neglect the backreactions or fluctuations of geometry at this level.

Since we want to study the general aspects that do not depend on any specific braneworld model we will consider a generic background metric given by
\begin{equation}
ds^{2}\!=\!g_{MN}dX^{M}dX^{N}\!=\!e^{2\sigma\left(y\right)}\!\left(\eta_{\mu\nu}dx^{\mu}dx^{\nu}\!+\!\eta_{jk}dy^{j}dy^{k}\right), \label{Sec-3-01}
\end{equation}
where $\eta_{\mu\nu}=\mbox{diag}\left(-1,1,1,1,...\right)$, $\eta_{jk}=\delta_{jk}$ (Kronecker Delta) and the warp factor $\sigma\!\left(y\right)$ depends only on the transverse extra dimensions $y^{j}$. Throughout the manuscript, capital indexes $M,N$ assume value on all $D$ dimensions; greek indexes are related to brane coordinates and run from $\mu,\nu=(1,2,...,D-d)$ and latin indexes are related to the $d$ extra dimensions and run from $j,k=(D-d+1,D-d+2,...,D)$.

At this point is important to note that the action (\ref{Sec-3-02}) is invariant by general coordinate transformation. Thus, when we perform a Lorentz transformation at the brane,
\begin{equation}
A^{M'}=L^{M'}_{\ M}A^{M}\ \ \ \rightarrow\ \ \ \left(\begin{matrix}
A^{\mu'} \\ 
B^{j'}
\end{matrix} \right)=\left(\begin{matrix}
\Lambda^{\mu'}_{\ \mu}& 0 \\ 
0 & \delta^{j'}_{\ j}
\end{matrix}\right)\left(\begin{matrix}
A^{\mu} \\ 
B^{j}
\end{matrix} \right),\label{Lorentz}
\end{equation}
where $\Lambda^{\mu'}_{\ \mu}$ is an usual Lorentz transformation in Minkowski spacetime. It makes clear that the components $A^{\mu}$ will be a  Lorentz vector at the brane. Also, the components  $B^{j}$ will be Lorentz scalars at the brane. Therefore, it is convenient to split the analysis of localization for this two fields (sectors).
Another important point is that the above action is not gauge invariant. Nevertheless, this is not a problem since an effective gauge theory can still be obtained on the brane and it contains a gauge field with all desired properties: massless, gauge and Lorentz invariant. 

Finally, in this and the next sections we will need of explicit expressions for the scalar and the Ricci tensor obtained from the metric (\ref{Sec-3-01}). The Ricci tensor is given by
\begin{eqnarray}
R_{MN}\!=\!-\!\left[\left(D\!-\!2\right)\delta^{k}_{M}\delta^{j}_{N}\!+\!\eta_{MN}\eta^{kj}\right]\!\partial_{k}\partial_{j}\sigma\!(y)
\ \ \ \ \ \ \ \ \ \ \ \ \ \ \ \ \ \nonumber\\+\!\left(D\!-\!2\right)\!\left[\delta^{k}_{M}\delta^{j}_{N}-\eta_{MN}\eta^{kj}\right]\!\partial_{k}\sigma\!(y) \partial_{j}\sigma\!(y);\label{Sec-3-03}
\end{eqnarray}
and the Ricci scalar by
\begin{eqnarray}
R =-\left(D-1\right)e^{-2\sigma\!(y)}\eta^{jk}\left[2\partial_{k}\partial_{j}\sigma\!(y)
+\left(D-2\right)\partial_{k}\sigma\!(y)\partial_{j}\sigma\!(y)\right].\label{Sec-3-04}
\end{eqnarray}
In the above equations and from now on we will use the definition $\partial_{k}\equiv\frac{\partial}{\partial y^{k}}$.

\subsection{\boldmath Localization of Transversal Sector of $U(1)$ Vector Field $-$ $\hat{\mathcal{A}}^{\mu}$}\label{Sec-3-1}
Due to the above discussion, and without loss of generality, it is convenient to split the $D$-dimensional vector field as $\mathcal{A}_{N}=\left(\hat{\mathcal{A}}_{\mu}+\partial_{\mu}\phi,\mathcal{B}_{k}\right)$, where $\hat{\mathcal{A}}_{\mu}$ is a transverse $(D-d)$-vector field on the brane defined such that $\eta^{\mu\nu}\partial_{\mu}\hat{\mathcal{A}}_{\nu}=0$. Note that the splitting of $\mathcal{A}_{N}$ field does not specify any gauge for the vector field, in fact the action (\ref{Sec-3-02}) is not gauge-invariant, therefore such symmetry can not be used to exclude degrees of freedom. 

With the above definitions the action (\ref{Sec-3-02}) can be split in two independent parts [see appendix (\ref{Apendice})]
\begin{equation}
 S=S_{\perp}[\hat{\mathcal{A}}_{\mu}]+S\left[\phi,\mathcal{B}_{k}\right],\label{Sec-3-05}
\end{equation}
where $S_{\perp}[\hat{\mathcal{A}}_{\mu}]$ contains only the transverse sector
\begin{eqnarray}
\! \! S_{\perp}=-\!\!\int \! d^{(D-d)}x d^{d}y \sqrt{-g}\left\lbrace\frac{1}{4}g^{\mu\nu}g^{\rho \lambda}\hat{\mathcal{F}}_{\mu\rho}\hat{\mathcal{F}}_{\nu\lambda}+\frac{1}{2}g^{\mu\nu}g^{jk}\partial_{j}\hat{\mathcal{A}}_{\mu}\partial_{k}\hat{\mathcal{A}}_{\nu}\right.\nonumber\\+\left.\frac{\lambda_{1}}{2}Rg^{\mu \nu}\hat{\mathcal{A}}_{\mu}\hat{\mathcal{A}}_{\nu}+\frac{\lambda_{2}}{2}g^{\mu\nu}g^{\rho \lambda}R_{\mu\rho}\hat{\mathcal{A}}_{\nu}\hat{\mathcal{A}}_{\lambda}\right\rbrace,\label{Sec-3-06}
\end{eqnarray}
and another part contains only the longitudinal and scalar components of the vector field, $\partial_{\mu}\phi$ and $B_{k}$ respectively. The explicit form of $S\left[\phi,B_{k}\right]$ was not written because it is not necessary for the discussion of this section. In next section, we will study the localization of $B_{k}$ components directly from (\ref{Sec-3-02}). From the action (\ref{Sec-3-06}) and using the metric (\ref{Sec-3-01}), we can obtain the following equation of motion for the transverse sector $\hat{\mathcal{A}}_{\mu}$,
\begin{eqnarray}
-e^{-(D-4)\sigma}\partial_{k}\left(e^{(D-4)\sigma}\partial^{k}\hat{\mathcal{A}}^{\lambda}\right)+\lambda_{1}Re^{2\sigma}\hat{\mathcal{A}}^{\lambda}+\lambda_{2}R^{\lambda}_{\ \ \mu}\hat{\mathcal{A}}^{\mu}=\partial_{\nu}\hat{\mathcal{F}}^{\nu\lambda},\label{Sec-3-07}
\end{eqnarray}
where the Minkowski metric was used to lower/raise indexes. 

In order to solve the equation (\ref{Sec-3-07}) we first need of explicit form of $R_{\mu\rho}$. From Eq. (\ref{Sec-3-03}) we can get that it is diagonal and given by
\begin{eqnarray}
R_{\mu\rho}\!=\!-\!\left[\partial_{k}\partial^{k}\sigma(y)\!+\!(D\!-\!2)\partial_{k}\sigma(y) \partial^{k}\sigma(y)\right]\! \eta_{\mu\rho}\!\equiv\!-h(y)\eta_{\mu\rho}.\label{Sec-3-03a}
\end{eqnarray}
With this and by performing the standard decomposition $\hat{\mathcal{A}}_{\nu}=A_{\mu}(x)\chi(y)e^{-\frac{(D-4)}{2}\sigma(y)}$, we find that Eq. (\ref{Sec-3-07}) reduces to the following equations. One for the transverse vector field $A_{\mu}(x)$ given by
\begin{eqnarray}
\partial_{\nu}F^{\nu\mu}(x)=m^{2}A^{\mu}(x)\label{Sec-3-08}
\end{eqnarray}
and another which drives the mass modes of the theory, given by
\begin{eqnarray}
-\partial_{k}\partial^{k}\chi\!+\!\left[\frac{(D\!-\!4)^{2}}{4}\partial_{k}\sigma\!\left(y\right)\!\partial^{k}\sigma\!\left(y\right)\!\right.&+&\!\frac{(D\!-\!4)}{2}\partial_{k}\partial^{k}\sigma\!\left(y\right)\!\nonumber\\&+&\left.\!\lambda_{1}Re^{2\sigma}\!-\!\lambda_{2}h\left(y\right)\right]\!\chi\!=\!m^{2}\chi.\label{Sec-3-09}
\end{eqnarray}
With this,  the action (\ref{Sec-3-06}) can be written as
\begin{eqnarray}
S_{\perp}=-\int d^{d}y\chi^{2}\int d^{D-d}x\left\lbrace\frac{1}{4}F_{\mu\rho}F^{\mu\rho}+\frac{1}{2}m^{2}A_{\nu}A^{\nu}\right\rbrace\label{Sec-3-10}
\end{eqnarray}
and a consistent effective theory localized over the brane requires that the integral
\begin{eqnarray}
K=\int d^{d}y\chi^{2}\label{Sec-3-10a}
\end{eqnarray}
be finite. The expression (\ref{Sec-3-09}) is a Schrödinger-like equation with potential given by 
\begin{equation}
 U\!(y^{j})=c_{1}\partial_{k}\partial^{k}\sigma\!\left(y\right)+c_{2}\partial_{k}\sigma\!\left(y\right)\partial^{k}\sigma\!\left(y\right), \label{Sec-3-09a}
 \end{equation} 
where we have used the relations (\ref{Sec-3-04}) and (\ref{Sec-3-03a}) to obtain the above result. Equation (\ref{Sec-3-09}) is very convenient because it allows us to obtain qualitative informations about the system just by the analysis of potential $U$. As known from quantum mechanics, the asymptotic behavior of $U$ determines whether the states $\chi(y)$ can be normalized or not. If $\lim_{|y|\rightarrow\infty} U\!(y)=\infty$, then we have a confining system and all states can be normalized; if otherwise $\lim_{|y|\rightarrow\infty} U\!(y)=-\infty$, then the states cannot be normalized; and finally, when $\lim_{|y|\rightarrow\infty} U\!(y)=u_{0}$ (constant) we have two possibilities: a) if $m^{2}<u_{0}$ the modes are normalized and b) if  $m^{2}=u_{0}$ the analysis must be performed case by case. As we do not know anything about the warp factor $\sigma\!\left(y\right)$, these questions can not be solved yet. Thus, let us obtain a solution for the zero-mode and carry out the discussion of localization.

\subsubsection{\boldmath Localization of Zero-Mode  $-$ $\hat{\mathcal{A}}_{0}^{\mu}$}\label{Sec-3-2}
Let us propose a zero-mode solution for equation (\ref{Sec-3-09}) as $\chi_{0}=e^{a\sigma(y)}$. It is easy to see that it will satisfy (\ref{Sec-3-09}) when $a=c_{1}$ and $a^{2}=c_{2}$. These conditions can be reduced to the condition $c_{1}^{2}=c_{2}$ and can always be satisfied by adjusting the coupling constants $\lambda_{1}$ and $\lambda_{2}$ in the action (\ref{Sec-3-06}). This possible values of  $\lambda_{1}$ and $\lambda_{2}$ will be analyzed latter. An important consequence of the condition $c_{1}^{2}=c_{2}$ is that it allows us to factor Eq. (\ref{Sec-3-09}) as  
\begin{equation}
\left[-\partial_{k}+c_{1}\partial_{k}\sigma(y)\right]\!\left[\partial^{k}+c_{1}\partial^{k}\sigma(y)\right]\chi(y^{j})=m^{2}\chi(y^{j}).\label{Sec-3-09b}
\end{equation}
This is analogous to a supersymmetric quantum mechanical problem, such that we can conclude that there are not tachyonic modes in the spectrum, \textit{i.e.}, $m^{2}\geq 0$. About the normalization of zero-mode $\chi_{0}(y)=e^{a\sigma(y)}$, we can infer from Eq. (\ref{Sec-3-10a}) that $\chi_{0}^{2}$ should go to zero faster than $\left|y^{j}\right|^{-d}$, when $\left|y^{j}\right|\rightarrow\infty$. This is the most we can obtain by considering an arbitrary $\sigma(y)$. 

In order to reach some conclusion about the localization of zero mode we will need of more information about our system. As said before, we can obtain information about normalization by analyzing the asymptotic behavior of $U$.  Note that our effective potential (\ref{Sec-3-09a}) depends only on the background warp factor $\sigma(y)$, so we must specify, at least, the behavior of  $\sigma(y)$ when $\left|y^{j}\right|\rightarrow\infty$. Since we look for RS-like models we will required an asymptotically $AdS$ background. This means that we must consider  $\lim_{|y^{j}|\to \infty}R=-\kappa$, with $\kappa > 0$ and constant. This leads to the following asymptotic behavior for $\sigma(y)$:
\begin{equation}
\lim_{|y^{j}|\to \infty}\sigma(y^{j})=-\ln\!\left(\sum_{j}\beta_{j}|y^{j}|\right),\label{Sec-3-09c}
\end{equation}
with $\beta_{j}$'s constants. With (\ref{Sec-3-09c}) we finally obtain from (\ref{Sec-3-09a}) that $\lim_{|y|\to \infty}U\!(y^{j})= 0$. Therefore the analysis of the potential is not conclusive about the localization of massless mode $(m^{2}=0)$ and we must analyze the solution of the equations of motion.
By plugging  (\ref{Sec-3-09c}) in our solution to the zero-mode, we can get the asymptotic behavior of our solution which is given by 
 \begin{equation}
 \lim_{|y^{j}|\to \infty}\chi^{2}_{0}(y^{j})=\left(\sum_{j}\beta_{j}|y^{j}|\right)^{-2a}.
 \end{equation}
If now we substitute this in the integral (\ref{Sec-3-10a}), we can conclude that it will be finite if we have $2a=2c_{1} > d$, where $d$ is the number of large transverse extra dimensions. Therefore we have two conditions given by 
\begin{equation}\label{conditions}
c_1^2=c_2\ \ \ \mbox{and}\ \ \ \ 2c_{1} > d
\end{equation}
which must be imposed to our system.

With the conditions (\ref{conditions}) we can analyze for which range of parameters $\lambda_{1}$ and $\lambda_{2}$ the massless mode ($U(1)$ gauge field) can be confined on the brane.  For this we use the relations (\ref{Sec-3-04}) and (\ref{Sec-3-03a}) to obtain explicit forms of $c_{1}$ and $c_{2}$, which are given by 
\begin{equation}
c_{1}=\frac{D-4}{2}-2\lambda_{1} \left(D-1\right)-\lambda_{2}\ \ \ \mbox{and}\ \ \ \ c_{2} =\frac{\left(D-4\right)^{2}}{4}-\left(\lambda_{1} \left(D-1\right)+\lambda_{2}\right)\left(D-2\right).\label{Sec-3-09d}
\end{equation}
Let us list some particular cases:
\begin{itemize}
\item[(i)] {\it Free gauge field} ($\lambda_{1}=\lambda_{2}=0$) - From Eq. (\ref{Sec-3-09d}) we obtain $2c_{1}=(D-4)$.  Thus, as we must have $2c_{1} > d$ in order to confine $\hat{\mathcal{A}}^{\mu}$, we get $(D-4)>d$. For example, if we consider a $3$-brane ($d=D-4$), we can conclude that the free gauge field cannot be localized for any number of extra dimensions.  

\item[(ii)] {\it Ricci scalar coupling} ($\lambda_{2}=0$) - When we use that $c^{2}_{1}=c_{2}$, we obtain that $\lambda_{1}=\frac{D-6}{4 \left(D-1\right)}$ and by using the localization condition $2c_{1} > d$, we obtain $2>d$. Therefore, the gauge field is localized only for models with co-dimension one. It is important to stress that for this and the next cases, \textit{Hodge duality} symmetry does not apply due to the presence of the interaction term.

\item[(iii)] {\it Ricci tensor coupling} ($\lambda_{1}=0$) - this allows us to write $2c_{1}=(D-4)-2\lambda_{2}>d$. From $c^{2}_{1}=c_{2}$ we get  $\lambda_{2}=-2$ and the localization of the transverse sector $\hat{\mathcal{A}}^{\mu}$ occurs for any number of extra dimensions. 

\item[(iv)] Finally, we have the case where $\lambda_ {1}$ and $\lambda_ {2}$ are non-zero. For this situation, there is a sub case where we can impose $\lambda_ {1}=-\frac{1}{2}\lambda_{2}$. This constraint allows us to combine the interaction terms with the scalar and the Ricci tensor in action (\ref{Sec-3-06}) to generate the Einstein tensor. This particular coupling gives us the same conclusions of the case (i). For the general situation, we obtain from $c^{2}_{1}=c_{2}$ the relation
\begin{equation}
\lambda^{(\pm)}_ {2}=-2\lambda_ {1}(D-1)-1\pm\sqrt{\lambda_ {1}(D-1)(D-2)+1}\label{lambda}
\end{equation}
and consequently
\begin{equation}
c^{(\pm)}_{1}=\frac{D-2}{2}\pm\sqrt{\lambda_ {1}(D-1)(D-2)+1}.
\end{equation} 
By imposing $2c_1>d$ we get 
\begin{equation}
\lambda_{1}\geq-\frac{1}{(D-1)(D-2)},\ \ \mbox{for}\ \ c^{(+)}_{1}
\end{equation} 
and 
\begin{equation}
-\frac{1}{(D\!-\!1)(D\!-\!2)}\leq\lambda_1<-\frac{4\!-\!\left(D\!-\!d\!-\!2\right)^{2}}{4(D\!-\!1)(D\!-\!2)},\ \ \mbox{for}\ \ c^{(-)}_{1}.
\end{equation} 
Therefore the localization can always be accomplished by just imposing that $\lambda_ {1}$ satisfy the above conditions. For example, if we choose $d=D-4$ we obtain that $\lambda_ {1}$ must belongs to the range $[-\frac{1}{(D-1)(D-2)},\infty)$ for $c^{(+)}_{1}$ or $[-\frac{1}{(D-1)(D-2)},0)$ for $c^{(-)}_{1}$. We should point to the fact that in this case the parameters $\lambda_1,\lambda_2$ are not completely fixed. This will be important when we consider the localization of scalar components of the vector field in the next section.
 \end{itemize}
The above discussion reveals some important points about the gauge field localization. Among the main results, we stress that the background is completely generic. We did not need to specify whether the model consists of a delta or a thick-like brane. Thus, this localization mechanism does not suffer from the `problem' of being sensitive to the thickness of the brane. Another important point is that the existence of a zero-mode solution for the gauge field imply that the model has no tachyon. These are the reasons why the geometric coupling has a general validity, allowing the confinement of $U(1)$ gauge field for an wide variety of braneworld models.

To conclude the discussion about transverse gauge field, we will carry out a comparative analysis between our zero-mode $\chi^{(\mbox{gauge})}(y)=\exp\left[c_{1}\sigma(y)\right]$ and the one found for the gravity field in Ref. \cite{Csaba}. Since the consistence of gravity is the starting point of braneworld models, it is interesting to analyze if its localization imply the localization of the gauge field. In ref. \cite{Csaba}, Csaba Csáki {\it et.al.}  performed a study about the universal aspects of gravity localization in braneworld models. From the standard Einstein-Hilbert  action, the authors studied the metric fluctuations on the background like that of Eq. (\ref{Sec-3-01}). An interesting point is that the zero-mode found for the gravitational field is given by $\psi^{(\mbox{gravity})}(y)=\exp\left[(\frac{D-2}{2})\sigma(y)\right]$. By imposing that this solution is square integrable they reach a general condition for gravity localization given by $D-2>d$.  In the case of the gauge field, in Eq. (\ref{conditions}) we also found a similar condition given by $2c_{1} > d$. In this way, if $2c_{1} \geq (D-2)$,  the localization of gravitational field is enough to ensure the localization of gauge field. Looking at the cases (i-iv) above discussed, only the configurations (iii) and (iv) can satisfy $2c_{1} \geq (D-2)$. For the case (iii) $2c_{1}=D\geq(D-2)$, therefore we get that the localization of gravity implies the localization of the gauge field.  For the case (iv) we see that only the solution $c^{(+)}_{1}$ can satisfy the relation $2c_{1} \geq (D-2)$. In Figs. (\ref{figura2a}) and (\ref{figura2b}) we give a plot of values of $\lambda_{1}$ and $\lambda_{2}$, as a function of the dimension $D$, which allow the localization of gauge field for some cases discussed above.  

\newpage
\subsection{\boldmath Localization of the Scalar Components $-$ $\mathcal{B}^{j}$}\label{Sec-3-3}

It is common in the literature to consider the trivial solution for the fields $\mathcal{B}_{j}$. The reason is that if these fields are not localized, its backreaction could modify the background \cite{Oda3}. Despite of this, in this section we must consider the more general case and study the possibility of localizing the components  $\mathcal{B}_{j}$. We will emphasize on the possibility that these fields can be confined simultaneously with the transverse gauge field $\hat{\mathcal{A}}^{\mu}$. We must point that the localization of these components have no consequence over the Coulomb law. Actually, by Eq. (\ref{Lorentz}), an observer at the brane sees these components as {\it scalar fields} and these have no relation to Coulomb's scalar potential. However, the effective theory for this fields over the brane could be interpreted as Higgs fields or maybe as dark energy, for example.

In order to study the localization of the scalar components $\mathcal{B}_{j}$, we will carry out the variation of action (\ref{Sec-3-02}) with respect to $\mathcal{A}^{M}$. By doing this, we get the following equation of motion
\begin{equation}
\frac{1}{\sqrt{-g}}\partial_{N}\left(\sqrt{-g}g^{MN}g^{PQ}\mathcal{F}_{MP}\right)=\lambda_{1} Rg^{MQ}\mathcal{A}_{M}+\lambda_{2} g^{MQ}g^{NP}R_{MN}\mathcal{A}_{P},\label{Sec-3-3-01}
\end{equation}
and due to the anti-symmetry of $\mathcal{F}_{MP}$, we also obtain the constraint
\begin{equation}
\partial_{Q}\left(\lambda_{1} R\sqrt{-g}g^{MQ}\mathcal{A}_{M}+\lambda_{2} \sqrt{-g}g^{MQ}g^{NP}R_{MN}\mathcal{A}_{P}\right)=0.\label{Sec-3-3-02}
\end{equation}
From these expressions, the equation of motion for the scalar fields $\mathcal{B}^{j}$ can be obtained. For this we choose $Q=k$ in the Eq. (\ref{Sec-3-3-01}) to obtain
\begin{eqnarray}
\partial_{\mu}\partial^{\mu}\mathcal{B}^{j}+e^{-(D-4)\sigma}\partial_{k}\!\left(e^{(D-4)\sigma}\mathcal{B}^{kj}\right)\ \ \ \ \ \ \ \ \ \ \ \ \ \ \ \ \ \ \ \ \ \ \ \ \nonumber\\-\left(\lambda_{1} R\delta^{j}_{\ k}+\lambda_{2} e^{-2\sigma}R^{j}_{\ k}\right)\!e^{2\sigma}\!\mathcal{B}^{k}=\partial^{j}\partial_{\mu}\mathcal{A}^{\mu}.\label{Sec-3-3-03}
\end{eqnarray}
We also write the constraint (\ref{Sec-3-3-02}) as 
\begin{eqnarray}
e^{-(D-2)\sigma}\partial_{k}\left[e^{(D-2)\sigma}\left(\lambda_{1} R\delta^{j}_{\ k}+\lambda_{2}e^{-2\sigma}R^{j}_{\ k}\right)\!\mathcal{B}^{k}\right]\ \ \ \ \ \ \ \ \ \ \ \ \ \ \ \ \ \ \ \ \ \ \ \ \nonumber\\+\left[\lambda_{1} R-\lambda_{2} e^{-2\sigma}h(y)\right]\partial_{\mu}\mathcal{A}^{\mu}=0.\label{Sec-3-3-03a}
\end{eqnarray}
In the above equations $\mathcal{B}^{kj}=\partial^{k}\mathcal{B}^{j}-\partial^{j}\mathcal{B}^{k}$ and the indexes are lowered/raised using the Minkowski metric (remember that $\eta_{jk}=\delta_{jk}$). Looking at the equations (\ref{Sec-3-3-03}) and (\ref{Sec-3-3-03a}), we realize that the solution for the fields $\mathcal{B}^{j}$ cannot be obtained as directly as for the field $\hat{\mathcal{A}}^{\mu}$. Actually, since the equations (\ref{Sec-3-3-03}) are coupled, its complete solution is model-dependent, {\it i.e.}, the warp factor must be known. Despite of this, some general properties of the system can be obtained by studying the asymptotic behavior of Eqs. (\ref{Sec-3-3-03}) and (\ref{Sec-3-3-03a}).

From now on we will look for a solution of  Eqs. (\ref{Sec-3-3-03}) and (\ref{Sec-3-3-03a}) when $|y^{j}|\to \infty$. In this way, we can study under what conditions they are convergent and square integrable. In this limit, the asymptotic warp factor (\ref{Sec-3-09c}) can be used and thus we get some simplifications on our system. First we see from Eq. (\ref{Sec-3-03}) that  $R_ {jk}$ can be written as
\begin{eqnarray}
R_{jk}&=&\left(D-2\right)\left[-\partial_{j}\partial_{k}\sigma\left(y\right)+\partial_{j}\sigma\left(y\right)\partial_{k}\sigma\left(y\right)\right]-h\left(y\right)\eta_{jk}\nonumber\\ &\equiv &\left(D-2\right)\Omega_{jk}-h\left(y\right)\eta_{jk},\label{Sec-3-3-04}
\end{eqnarray}
with $h(y)$ defined in Eq. (\ref{Sec-3-03a}). By using the asymptotic warp factor (\ref{Sec-3-09c}) we can find
\begin{eqnarray}
\Omega_{jm} =-\partial_{m}\partial_{j}\sigma\left(y\right)+\partial_{m}\sigma\left(y\right)\partial_{j}\sigma\left(y\right)\rightarrow 0,\label{Sec-3-3-06}
\end{eqnarray}
and thus (\ref{Sec-3-3-04}) gets the form
\begin{eqnarray}
R_{jk}\to h\left(y\right)\eta_{jk}.\label{Sec-3-3-04a}
\end{eqnarray}
Another simplification is that $\lambda_{1}R-\lambda_{2}h(y)e^{-2\sigma}\equiv g\left(y\right) $ goes to a constant value $g\left(y\right) \to C_{1}$. With these simplifications the constraint (\ref{Sec-3-3-03a}) simplifies to 
\begin{equation}
e^{-(D-2)\sigma}\partial_{k}\left[e^{(D-2)\sigma}\mathcal{B}^{k}\right]+\partial_{\mu}\mathcal{A}^{\mu}=0.
\end{equation}
Substituting this in Eq. (\ref{Sec-3-3-03}) we finally get  
\begin{eqnarray}
e^{-\left(D-4\right)\sigma}\partial_{k}\!\left[e^{\left(D-4\right)\sigma}\partial^{k}\mathcal{B}^{j}\right]\!+\!\partial_{\rho}\partial^{\rho}\mathcal{B}^{j}\!-\!g\!\left(y\right)\!e^{2\sigma}\!\mathcal{B}^{j}\ \ \ \ \ \ \ \ \ \ \ \ \ \ \ \ \ \ \ \ \ \ \ \ \nonumber\\+2\partial^{j}\!\mathcal{B}^{k}\partial_{k}\sigma +\!\left(D-2\right)\mathcal{B}^{k}\partial^{j}\partial_{k}\sigma=0.\label{Sec-3-3-07}
\end{eqnarray}
The above equation is the main one which we will try to solve. In this way, we intend, at least, to get the some conditions for which the scalar components can be localized on the brane.

Let us propose the transformation $\mathcal{B}^{k}(x,y)=B^{k}(x,y)e^{-\frac{\left(D-2\right)}{2}\sigma}$ in Eq. (\ref{Sec-3-3-07}). With this, the last two terms can be eliminated and we obtain
\begin{eqnarray}
\partial_{k}\partial^{k}B^{j}(x,y)&-&\left[\frac{\left(D-2\right)}{2}\partial_{k}\partial^{k}\sigma +\frac{\left(D-2\right)\left(D-6\right)}{4}\partial_{k}\sigma\partial^{k}\sigma+g\!\left(y\right)\!e^{2\sigma}\right] B^{j}(x,y)\nonumber\\&+&\partial_{\rho}\partial^{\rho}B^{j}(x,y)+2\left[\partial^{j}B^{k}(x,y)-\partial^{k}B^{j}(x,y)\right]\partial_{k}\sigma=0.\label{Sec-3-3-08}
\end{eqnarray}
Note that the above equation has a very convenient form. We can carry out a contraction of Eq. (\ref{Sec-3-3-08}) with $\partial_{j}\sigma$ to obtain an equation that can be easily solved. By doing this and using the asymptotic warp factor (\ref{Sec-3-09c}) we obtain 
\begin{eqnarray}
\left[\frac{\left(D\!-\!2\right)}{2}\left(\partial_{k}\partial^{k}\sigma\!+\!\frac{\left(D\!-\!6\right)}{2}\partial_{k}\sigma\partial^{k}\sigma\right)\!+\!g\!\left(y\right)\!e^{2\sigma}\right]\!\Phi(x,y)\nonumber\\-\partial_{k}\partial^{k}\Phi(x,y)=\!\partial_{\rho}\partial^{\rho}\Phi(x,y),\label{Sec-3-3-09}
\end{eqnarray}
where we have defined 
\begin{equation}\label{Sec-3-3-09a}
\sum_{j}\mbox{sgn}(y^{j})\beta_{j}B^{j}(x,y)\equiv \Phi(x,y)
\end{equation} 
with $\mbox{sgn}(y^{j})$ the sign function defined in standard way. This definition can be used in  Eq. (\ref{Sec-3-3-08}) in order to remove the coupled terms. By doing this, we get a non-homogeneous source-like term as follows
\begin{eqnarray}
\partial_{k}\partial^{k}B^{j}(x,y)&-&\left[\frac{\left(D-2\right)}{2}\partial_{k}\partial^{k}\sigma +\frac{\left(D-2\right)\left(D-6\right)}{4}\partial_{k}\sigma\partial^{k}\sigma+g\!\left(y\right)\right] B^{j}(x,y)\nonumber\\&+&\partial_{\rho}\partial^{\rho}B^{j}(x,y)-2\partial^{k}B^{j}(x,y)\partial_{k}\sigma =2e^{\sigma}\partial^{j}\Phi(x,y).\label{Sec-3-3-13}
\end{eqnarray}
Note that the source of Eq. (\ref{Sec-3-3-13}) is given by the solution of Eq. (\ref{Sec-3-3-09}). Therefore, in order to solve (\ref{Sec-3-3-13}) we first need to solve Eq. (\ref{Sec-3-3-09}). Before this we can use the known fact of partial differential equation that the most general solution is given by the sum $B^{j}(x, y) = \tilde{B}^{j}(x, y) +\tilde{B}_p^{j}(x, y)$, where $\tilde{B}^{j}(x, y)$ is a solution to the homogeneous equation
\begin{eqnarray}
\partial_{k}\partial^{k}\tilde{B}^{j}(x,y)&-&\left[\frac{\left(D-2\right)}{2}\partial_{k}\partial^{k}\sigma +\frac{\left(D-2\right)\left(D-6\right)}{4}\partial_{k}\sigma\partial^{k}\sigma+g\!\left(y\right)\right]\tilde{B}^{j}(x,y)\nonumber\\&+&\partial_{\rho}\partial^{\rho}\tilde{B}^{j}(x,y)-2\partial^{k}\tilde{B}^{j}(x,y)\partial_{k}\sigma =0;\label{Sec-3-3a-18}
\end{eqnarray}
and  $\tilde{B}_p^{j}(x, y)$ is a particular solution of the complete equation
\begin{eqnarray}
\partial_{k}\partial^{k}\tilde{B}_p^{j}(x, y)&-&\left[\frac{\left(D-2\right)}{2}\partial_{k}\partial^{k}\sigma +\frac{\left(D-2\right)\left(D-6\right)}{4}\partial_{k}\sigma\partial^{k}\sigma+g\!\left(y\right)\right] \tilde{B}_p^{j}(x, y)\nonumber\\&+&\Box\tilde{B}_p^{j}(x, y)-2\partial^{k}\tilde{B}_p^{j}(x, y)\partial_{k}\sigma =2e^{\sigma}\partial^{j}\Phi(x,y);\label{Sec-3-3a-20aaa}
\end{eqnarray}

In this way, Eq. (\ref{Sec-3-3-08}) is reduced to the above system of equations (\ref{Sec-3-3-09}), (\ref{Sec-3-3a-18}) and (\ref{Sec-3-3a-20aaa}). The next step is to perform a separation of variables of these equations. First let us propose $\Phi(x, y) = \theta(x)\zeta(y)$, thus we get the two equations
\begin{eqnarray}
\square_{x}\theta(x)= M^{2}\theta(x);\label{Sec-3-3-11}
\end{eqnarray}
and
\begin{eqnarray}
-\partial_{k}\partial^{k}\zeta(y) +U(y)\zeta(y)\!=\!M^{2}\zeta(y),\label{Sec-3-3-12}
\end{eqnarray}
where
\begin{equation}\label{Sec-3-3-12a}
U(y)=\frac{\left(D\!-\!2\right)}{2}\!\left(\!\partial_{k}\partial^{k}\sigma\!+\!\frac{\left(D\!-\!6\right)}{2}\partial_{k}\sigma\partial^{k}\sigma\!\right)\!+\!g\!\left(y\right)\!e^{2\sigma}.
\end{equation}

If now we substitute $\Phi(x, y) = \theta(x)\zeta(y)$ in Eq. (\ref{Sec-3-3a-20aaa}) we see that the only way to separate the variables is by choosing $\tilde{B}_p^{j}(x, y)=\theta(x)Z^{j}(y)$. Therefore, the final form of the non-homogeneous equation is given by
\begin{eqnarray}
\partial_{k}\partial^{k}Z^{j}(y)&-&\left[\frac{\left(D-2\right)}{2}\partial_{k}\partial^{k}\sigma +\frac{\left(D-2\right)\left(D-6\right)}{4}\partial_{k}\sigma\partial^{k}\sigma+g\!\left(y\right)\right] Z^{j}(y)\nonumber\\&+&M^2Z^{j}(y)-2\partial^{k}Z^{j}(y)\partial_{k}\sigma =2e^{\sigma}\partial^{j}\zeta(y).\label{Sec-3-3a-20}
\end{eqnarray}
Thus our final solution is in the form
\begin{equation}
B^{j}(x, y) = \tilde{B}^{j}(x, y) +\theta(x)Z^{j}(y)
\end{equation}
where, as said before, $\tilde{B}^{j}(x, y)$ is a solution of Eq. (\ref{Sec-3-3a-18}), $\theta(x)$ the solution of Eq. (\ref{Sec-3-3-11}) and $Z^{j}(y)$ the solution of Eq. (\ref{Sec-3-3a-20}). At this moment, the issue of zero-mode localization for the fields $ B^{j}(x, y)$ can be attained.

\subsubsection{\boldmath Localization of Zero-Mode $-$ $\mathcal{B}_{0}^{j}$}\label{Sec-3-4}
Before to study the zero-mode localization for the scalar components, we should point that from definition (\ref{Sec-3-3-09a}) and by using the Eq. (\ref{Sec-3-3-11}), we see that 
\begin{equation}
\square_{x}\Phi(x,y)= M^{2}\Phi(x,y)= \sum_{j}\mbox{sgn}(y^{j})\beta_{j}\square_{x} B^{j}(x,y).
\end{equation} 
If we are studying the zero-mode $(M^{2}=0)$ of the field $\Phi(x,y)$, then we should have $\sum_{j}\mbox{sgn}(y^{j})\beta_{j}\square_{x} B^{j}(x,y)=0$. With this, we  conclude that $\square_{x} B^{j}(x, y)=0 $ for each field $B^{j}(x, y)$ independently. Therefore, the massless mode analysis of  $\Phi(x,y)$ corresponds to the massless mode analysis of $B^{j}(x, y)$.

Now we can turn to the zero-mode solution of the fields $B^{j}(x, y)$. First, in view of the Eq. (\ref{Sec-3-3a-20}), we need to solve Eq. (\ref{Sec-3-3-12}) for the zero-mode $\zeta_{0}(y)$. By proposing the ansatz $\zeta_{0}(y)=e^{\tilde{b}\sigma(y)}$ for Eq. (\ref{Sec-3-3-12}) with $M^{2}=0$, we see that this is a solution when 
\begin{equation}\label{b}
\tilde{b}_{\pm}\!=\!-\frac{1}{2}\pm\frac{1}{2}\!\left[(D\!-\!3)^{2}\!-\!4(D\!-\!1)(\lambda_{2}\!+\!\lambda_{1}D).\right]^{\frac{1}{2}}
\end{equation}
Note that Eq. (\ref{Sec-3-3-12}) is a Schrödinger-like equation with potential $U$ given in Eq. (\ref{Sec-3-3-12a}). The potential $U$ is even by spatial inversion $(y^{j}\rightarrow -y^{j})$, and this generates a solutions for the field $\Phi(x, y)$ with well-defined parity (even or odd). We saw earlier that  the warp factor (\ref{Sec-3-09c}) is an even function of $y^{j}$. Therefore, since our solution $\zeta_{0}(y)$ is a power of the warp factor, we find that the field $\Phi_{0}(x,y)$ is even by the exchange $y^{j} \to -y^{j}$.

With $\zeta_{0}(y)$ obtained above, we can now look for a solution of the non-homogeneous equation (\ref{Sec-3-3a-20}), which becomes
\begin{eqnarray}
\partial_{k}\partial^{k}Z_{0}^{j}(y)&-&\left[\frac{\left(D-2\right)}{2}\partial_{k}\partial^{k}\sigma +\frac{\left(D-2\right)\left(D-6\right)}{4}\partial_{k}\sigma\partial^{k}\sigma+g\!\left(y\right)\right]Z_{0}^{j}(y)\nonumber\\&-&2\partial^{k}Z_{0}^{j}(y)\partial_{k}\sigma =2\tilde{b}e^{(\tilde{b}+1)\sigma(y)}\partial^{j}\sigma(y).\label{Sec-3-3-15}
\end{eqnarray}
To solve the above equation first note that,  as discussed previously, the field $\Phi_0(x, y)$ has even parity. Since we also have that $\Phi(x,y)=\sum_{j}\mbox{sgn}(y^{j})\beta_{j}B^{j}(x,y)$, $\mbox{sgn}(y^{j})B_0^{j}(x,y)$ must also be even.  This implies that $Z^{j}_{0}(y)$ must be odd. With this it is easy to check that the solution for (\ref{Sec-3-3-15}) is obtained from the ansatz $Z^{j}_{0}(y)= \mbox{sgn}(y^{j})|y^{j}|e^{(1+\tilde{b})\sigma(y)}$ (without sum in $j$). 

About the solution for the homogeneous Eq. (\ref{Sec-3-3a-18}), this can be found by performing a very similar treatment to that used in section  (\ref{Sec-3-2}). By doing this, we find
\begin{equation}
\tilde{B}^{j}(x, y)=\mbox{sgn}(y^{j})e^{(\tilde{b}+1)\sigma}\bar{B}_{0}^{j}(x).
\end{equation}
Finally, we arrive at the final solution for our system, given by
\begin{eqnarray}
\mathcal{B}_{0}^{j}(x,y)&=&e^{-\frac{(D-2)}{2}\sigma}B_{0}^{j}(x,y)\nonumber \\ &=&\mbox{sgn}(y^{j})e^{\left[\tilde{b}-\frac{(D-4)}{2}\right]\sigma}\left[\bar{B}_{0}^{j}(x)+\theta_{0}(x)|y^{j}|\right]\equiv\mbox{sgn}(y^{j})f^{j}(x,|y|).\label{Sec-3-3a-14a}
\end{eqnarray}
Beyond that, by definition (\ref{Sec-3-3-09a}) and due to the form of the warp factor (\ref{Sec-3-09c}), the fields $B_{0}^{j}(x,y)$ must satisfy   
\begin{equation}\label{Sec-3-3aa-14a}
\sum_{j}\beta_{j}\bar{B}_{0}^{j}(x)=0.
\end{equation} 
Eq. (\ref{Sec-3-3aa-14a}) gives us a constraint on the fields $\bar{B}_{0}^{j}(x)$, which at the end leads to the correct number of degrees of freedom. After all this, we can analyze under what conditions the zero-mode of the scalar components can be confined on the brane.

\begin{figure}[ht]
\centering
\includegraphics[width=.47\linewidth]{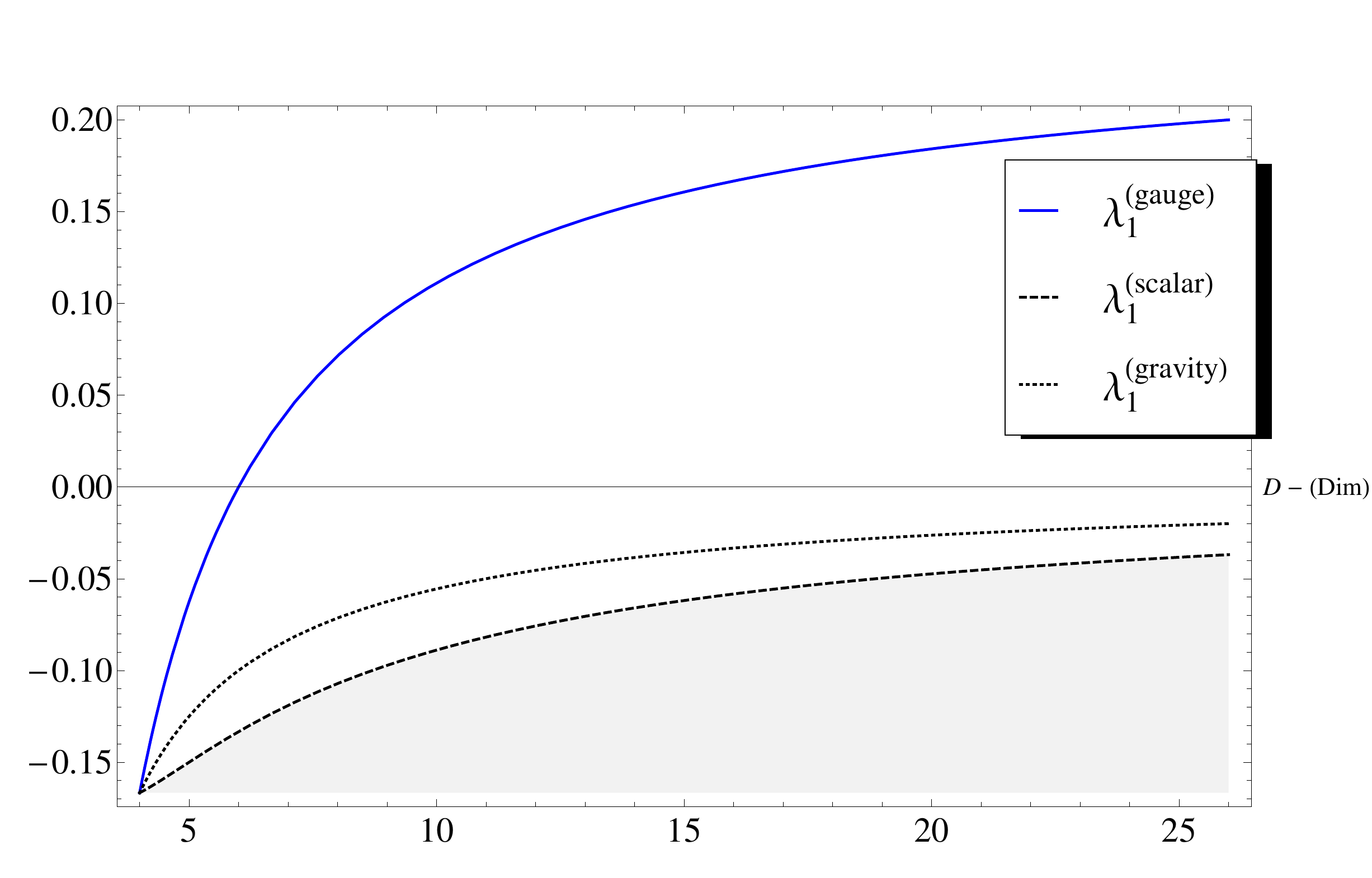}
\caption{Case (ii) of the sections (\ref{Sec-3-2}) and (\ref{Sec-3-4}) - The blue solid line gives the values of $\lambda_{1}\!(D)$ which allow the localization of the $\hat{\mathcal{A}}^{\mu}$ field. The region bellow the dashed line shows the values which allow the localization of the $\mathcal{B}^{j}$ fields. Also, the region bellow the dotted line are the values of $\lambda_{1}\!(D)$ for which the localization of gravity ensures the localization of $\hat{\mathcal{A}}^{\mu}$.}\label{figura2a}
\end{figure}

\begin{figure}[ht]
\centering
\includegraphics[width=.47\linewidth]{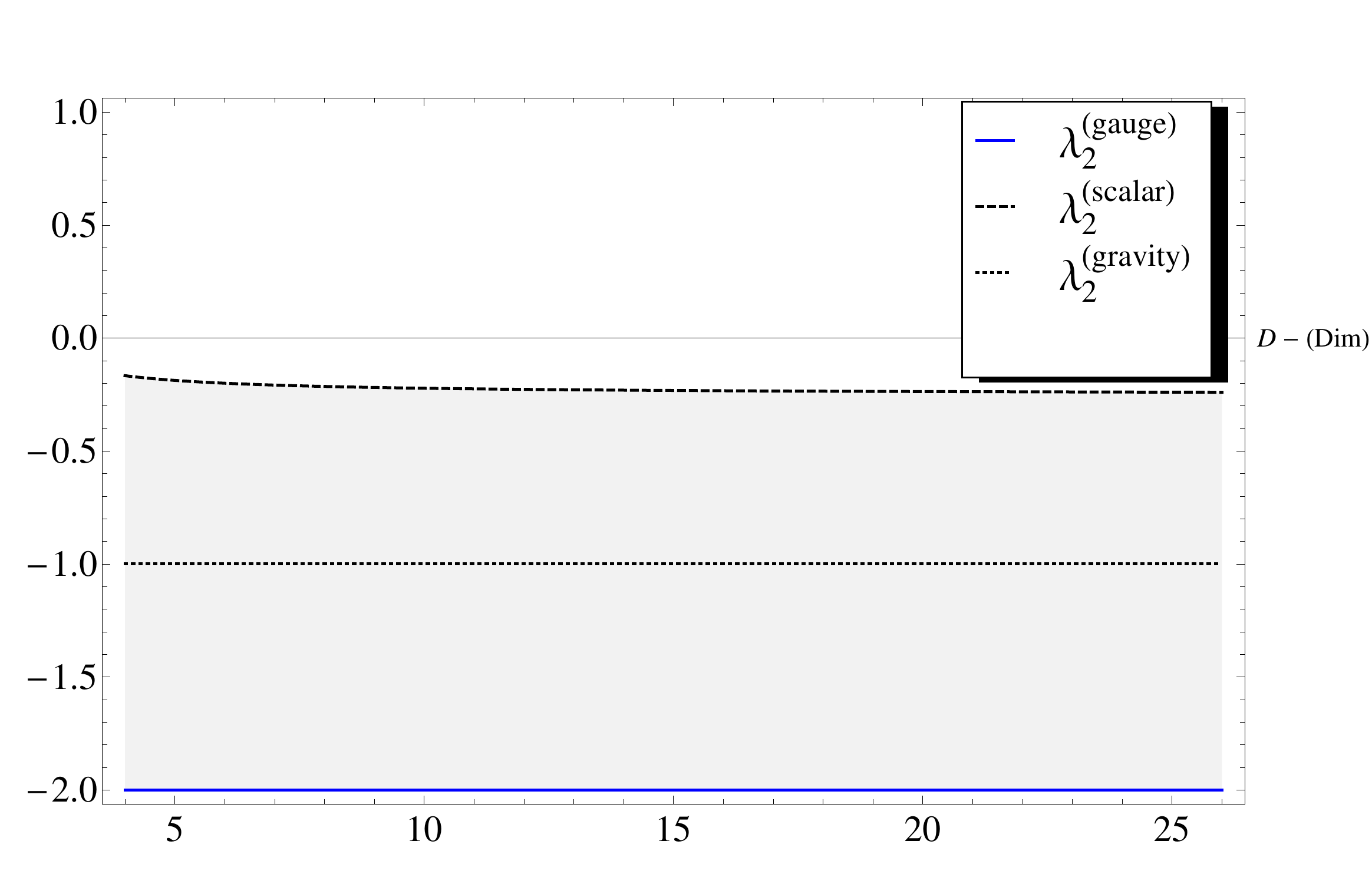}
\caption{Case (iii) of the sections (\ref{Sec-3-2}) and (\ref{Sec-3-4}) - The blue solid line gives the values of $\lambda_{2}\!(D)$ which allow the localization of the $\hat{\mathcal{A}}^{\mu}$ field. The region bellow the dashed line shows the values which allow the localization of the $\mathcal{B}^{j}$ fields. Also, the region bellow the dotted line are the values of $\lambda_{2}\!(D)$ for which the localization of gravity ensures the localization of $\hat{\mathcal{A}}^{\mu}$.}\label{figura2b}
\end{figure}

In the appendix (\ref{Apendice}) we split the action (\ref{Sec-3-02}) in two sectors given by Eq. (\ref{A-07}) for the gauge field $\hat{\mathcal{A}}^{\mu}$ and Eq. (\ref{A-08}) that contain the scalar fields  $\mathcal{B}^{j}$. In this last expression, the kinetic term that we must use to deal with the localization of massless mode is given by
\begin{eqnarray}
S_{0}[\mathcal{B}^{j}]=-\frac{1}{2}\int d^{(D-d)}x d^{d}y \sqrt{-g}g^{\mu\nu}g^{jk}\partial_{\mu}\mathcal{B}_{j}\partial_{\nu}\mathcal{B}_{k}.\label{Sec-3-3-18}
\end{eqnarray}
We are interested only in the convergence of the integral when $|y^{j}|\to\infty$. Thus, for large $|y^{j}|$, $\theta_{0}(x)|y^{j}|$ is the dominant term of our solution (\ref{Sec-3-3a-14a}). Therefore, if it is square integrable, all the solution will be. In our case, the $y^{j}$ integral in the action (\ref{Sec-3-3-18}) will be finite when $2\tilde{b}_{\pm}> d+2$. With this condition, we can analyze each of the cases bellow:
\begin{itemize}
\item[(i)] {\it Free scalar field} ($\lambda_{1}=\lambda_{2}=0$) - from the condition $2\tilde{b}_{\pm}>d+2$, we get $\pm(D-3)> d+3$. For the case $D=4+d$, for example,  it is easy to see that it is not possible to localize the scalar components.

\item[(ii)] {\it Ricci scalar coupling} ($\lambda_{2}=0$) - In this case, the convergence condition $2\tilde{b}_{\pm}> d+2$ leads to a constraint for $\lambda_{1}$ given by $ \lambda_{1}<-\frac{(d+3)^{2}-(D-3)^{2}}{4D(D-1)}$. With this, only the solution $\tilde{b}_{+}$ allows a zero-mode localized for the scalar components.  The Fig. (\ref{figura2a}) shows the values of $ \lambda_{1}(D)$  that allow the localization of scalar components over a $3$-brane ($D=d+4$). Remember that for the gauge field $\hat{\mathcal{A}}_{\mu}$, we obtained $\lambda_{1}=\frac{D-6}{4(D-1)}$. From the Fig. (\ref{figura2a}), we see that this value of $\lambda_{1}$ is not inside the region that allows the localization of two sectors simultaneously on the $3$-brane.

\item[(iii)] {\it Ricci tensor coupling} ($\lambda_{1}=0$) - For this, we get the constrain $ \lambda_{2}<-\frac{(d+3)^{2}-(D-3)^{2}}{4(D-1)} $ and, again, only the solution $\tilde{b}_{+}$ allows a zero-mode localized for the scalar components. For this case, the value $ \lambda_{2} = - 2$, defined in item (iii) of section (\ref{Sec-3-2}), obeys this relation for any number of transverse extra dimensions. In this way, there is the possibility of 'trapping' simultaneously both sectors on a $3$-brane with this kind of coupling. This can also be seen in Figure (\ref{figura2b}).

\item[(iv)] Finally, with $ \lambda_ {1}$ and $ \lambda_{2}$ nonzero, we get $\lambda_{1}D+\lambda_{2}<-\frac{(d+3)^{2}-(D-3)^{2}}{4(D-1)} $. Thus, we also can have both sectors localized simultaneously on a $3$-brane.
\end{itemize}
Of course, we only can ensure that the scalar components $\mathcal{B}^{j}$ are localized if we obtain the complete solution for them. Anyway, we get at least one general condition that must be imposed on our system such that we get confined massless scalar fields.  In the next section, we will apply this general results for a specific braneworld model.

\section{Application: Intersecting Brane Model}\label{Sec-5}
\begin{figure}[ht]
\centering
\includegraphics[width=.37\linewidth]{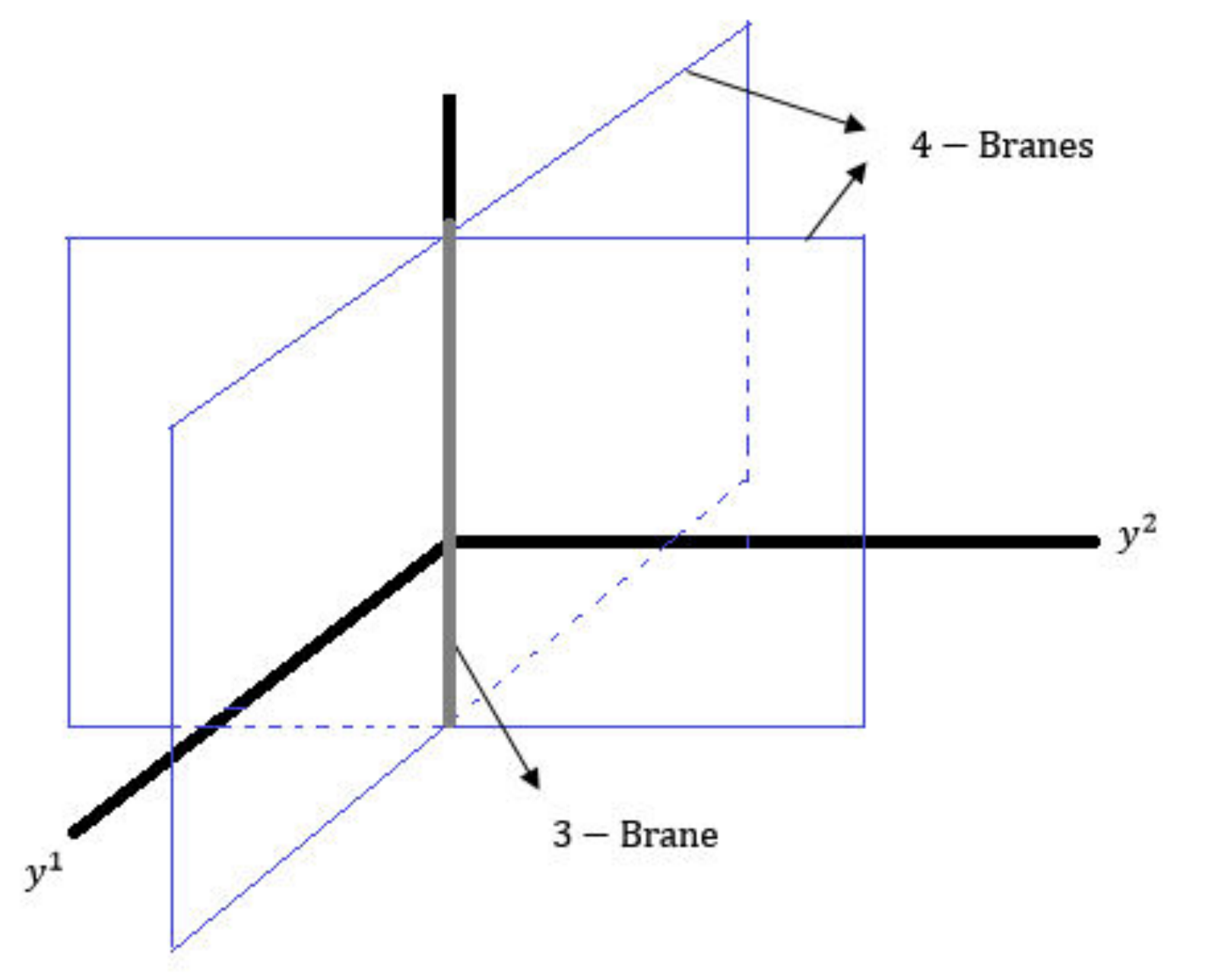}
\caption{Intersecting branes model with two transverse extra dimensions.}
\label{figura4}
\end{figure}

In section (\ref{Sec-3}), the general aspects of the $D$-vector field localization in a generic braneworld model was studied. Among the main general results, we obtained that the confinement  of this higher dimensional vector field generates an effective theory for one $U(1)$ gauge field $\hat{\mathcal{A}}^{\mu}$. Beyond this, there is an indicative that an effective scalar theory related to components  $\mathcal{B}^{j}$ can be also obtained. For the gauge field, it was found a zero-mode solution which is valid for any warp factor. However for the scalar components, it was only possible to find an asymptotic zero-mode solution. Here, we will apply the general results obtained in last section to a specific braneworld model and a more detailed discussion on the scalar components localization will be carried out. We will use the intersecting brane model developed by Arkani-Hamed {\it et. al.} \cite{Arkani}. In this model, the background metric is given by Eq. (\ref{Sec-3-01}) with the warp factor 
\begin{equation}
\sigma(y^{j})=-\ln\!\left(1+k\sum_{j}|y^{j}|\right).\label{Sec-5-01}
\end{equation}
The authors show that gravity is localized in a warped model generated by the intersection of  $d$ delta-like $(D-2)$-branes. In the figure (\ref{figura4}) it is shown an example with two transverse extra dimensions. For this case we have two delta-like $4$-branes intersecting one each other generating one $3$ dimensional brane. 

We will focus our attention on the discussion of the massless scalar components $\mathcal{B}_{j}$ of section (\ref{Sec-3-4}). This is because the solution for these components were obtained only in an asymptotic approximation, and it would be interesting to find a complete solution for these fields. In this way, we will be able to verify whether the results for scalar components can lead to a reasonable effective theory on the brane. Beyond this, from now on we will fix our brane with $3$ spatial dimensions, {\it i.e.}, $D=d+4$. As the branes in (\ref{Sec-5-01}) are delta-like, the zero-mode solutions $\mathcal{B}^{k}_{0}=B^{k}(x,y)e^{-\frac{\left(D-2\right)}{2}\sigma}$, with $B^{k}(x,y)$ given by Eq. (\ref{Sec-3-3a-14a}), must be valid for all $y^{j}\neq 0$. Thus, we only have to obtain the boundary conditions imposed by the branes on such solutions. An important point of this braneworld model is that each $(D-2)$-brane impose one boundary condition on the fields $\mathcal{B}_{j}$. Beyond this, at the intersection points we also must impose another boundary condition. Looking at the figure (\ref{figura4}), it means that each $4$-brane will introduce one boundary condition and the $3$-brane will impose another one.

As we saw, all the analysis for scalar fields was performed from equations (\ref{Sec-3-3-03}) and (\ref{Sec-3-3-03a}). For the present case, the warp factor (\ref{Sec-5-01}) allows us to write the following common factor of these equations
\begin{equation}
\lambda_{1} Re^{2\sigma}\delta_{\ j}^{k}+\lambda_{2} R^{k}_{\ j}=C_{0}e^{\sigma\!(y)}\delta^{k}_{\ j}\sum_{m}\delta(y^{m})-C_{1}e^{2\sigma}\delta^{k}_{\ j}+C_{2}e^{\sigma}\delta(y^{j})\delta_{\ j}^{k}.\label{Sec-5-03}
\end{equation}
In the above expression the constants $C_{0}$, $C_{1}$ and $C_{2}$ are given by
\begin{eqnarray}
C_{0}&=&2k\left[\lambda_{2}+2\lambda_{1}\left(D-1\right)\right];\nonumber \\
C_{1}&=&k^{2}\left(D-1\right)\left(D-4\right)\left(\lambda_{2}+\lambda_{1}D\right); \nonumber\\ 
C_{2}&=&2k\lambda_{2}\left(D-2\right).\label{Sec-5-04a}
\end{eqnarray}
From this we can carry out some considerations about the constraint (\ref{Sec-3-3-03a}). It can be written in the form
\begin{eqnarray}
e^{-(D-2)\sigma}\partial_{k}\!\!\left[\!e^{(D-2)\sigma}\!\!\left(\!\!C_{0}e^{-\sigma}\!\sum_{m}\!\delta(y^{m})-C_{1}+C_{2}e^{-\sigma}\delta(y^{k})\!\!\right)\!\!\mathcal{B}^{k}\right]\ \ \ \ \ \ \ \ \ \ \ \ \ \ \ \ \ \ \ \nonumber\\+\left[\!C_{0}e^{-\sigma}\!\sum_{m}\!\delta(y^{m})\!-\!C_{1}\!\right]\!\!\partial_{\mu}\mathcal{A}^{\mu}=0.\label{Sec-5-05}
\end{eqnarray}
Due to the presence of Dirac's delta, the following procedures can be performed. We will consider the above constraint for $y^{j}\neq 0$
\begin{equation}
\partial_{\mu}\mathcal{A}^{\mu}+e^{-(D-2)\sigma}\mbox{sgn}(y^{k})\partial_{k}\!\!\left[\!e^{(D-2)\sigma}f^{k}\right]=0,\label{Sec-5-06}
\end{equation}
where $f^{k}$ was defined in Eq. (\ref{Sec-3-3a-14a}). Next, we assume that it is valid for all $y^{j}$. In order to this to be valid we must substitute Eq. (\ref{Sec-5-06}) in Eq. (\ref{Sec-5-05}) such that it becomes
\begin{eqnarray}
2C_{0}\!\!\sum_{m\neq k}\!\delta(y^{m})e^{-\sigma}\!\sum_{k}\!f^{k}\delta(y^{k})-C_{0}k\!\!\sum_{m\neq k}\!\delta(y^{m})\!\!\sum_{k}f^{k}-2C_{1}\delta(y^{k})\!f^{k}=0.\label{Sec-5-07}
\end{eqnarray}
With this we can consider that we have solved our constraint and the above equation will impose boundary conditions on the fields $\mathcal{B}^{k}$. Beyond this boundary conditions, the fields $\mathcal{B}^{k}$ must also obey a set of boundary conditions obtained from equations of motion (\ref{Sec-3-3-03}). We saw that for this intersecting branes model each brane must impose a boundary condition on the fields $\mathcal{B}^{k}$. By using the figure (\ref{figura4}) as example, we will carry out one integration of Eqs. (\ref{Sec-3-3-03}) and (\ref{Sec-5-07}) into the range $[-\epsilon,\epsilon]$ for each extra coordinate keeping the others fixed, this will give us a set of boundary conditions. Next, we perform an integration in small volume containing the intersection point. This will give us another boundary conditions. For an arbitrary number of extra dimensions we must only extend this procedure.

By using the above procedure to Eq. (\ref{Sec-5-07}) we get the following boundary conditions
\begin{eqnarray}
C_{0}(D-5)\sum_{k}f^{k}(0)=0;\label{Sec-5-11}
\end{eqnarray}
\begin{equation}
2C_{1}f^{n}(y^{n}=0)-C_{0}k\sum_{k\neq n}f^{k}(y^{n}=0)=0.\label{Sec-5-12}
\end{equation}
Applying the same to Eq. (\ref{Sec-3-3-03}) we get  
\begin{eqnarray}
\left(C_{0}+C_{2}\right)\mathcal{B}_{0}^{k}(y^{k}=0,y^{j})\Big|_{y^{j}=c;\ j\neq k}=0;\label{Sec-5-08}
\end{eqnarray}
\begin{eqnarray}
C_{0}e^{\sigma\!(y)}\mathcal{B}_{0}^{k}(y^{m}=0,y^{j})\Big|_{y^{j}=c;\ m\neq j}=\left[\frac{d}{dy^{m}}\mathcal{B}_{0}^{k}(y^{m},y^{j})\right]^{y^{m}=\frac{\epsilon}{2}}_{y^{m}=-\frac{\epsilon}{2},\ y^{j}=c;\ m\neq j};\label{Sec-5-09}
\end{eqnarray}
\begin{eqnarray}
\frac{\left(d C_{0}+C_{2}\right)}{2}\mathcal{B}_{0}^{k}(0)=kc_{1}(d-1)\mathcal{B}_{0}^{k}(0)-kc_{1}\mbox{sgn}(y^{k}=0)\sum_{\substack{j\\j\neq k}}\mathcal{B}_{0}^{j}(0^{+}).\label{Sec-5-10}
\end{eqnarray}
The above equations are all the boundary conditions which must be imposed on the scalar fields. First, the equations (\ref{Sec-5-08}) and (\ref{Sec-5-10}) will be trivially satisfied if we define $\mbox{sgn}(0)=0$. Second, the condition (\ref{Sec-5-09}) imposes the constraint $C_{0}=-2k(\tilde{b}+1)+k (D-2)$ on the parameters of the theory. With this and by using Eqs. (\ref{b}) and (\ref{Sec-5-04a})  we get
\begin{equation}
\lambda^{(\pm)}_{2}=-2\lambda_{1}\left(D-1\right)-1\pm\sqrt{\lambda_{1}(D-1)(D-2)+1}.\label{Sec-5-14}
\end{equation}
The expression (\ref{Sec-5-14}) is exactly the relation obtained in Eq. (\ref{lambda}) for the localization of gauge field $\hat{\mathcal{A}}^{\mu}$. This is an excellent result, since we wish to confine both sectors (gauge and scalar fields) simultaneously. Finally, the equation (\ref{Sec-5-11}) gives us $\sum_{k}\bar{B}^{k}(x)=0$ and the Eq. (\ref{Sec-5-12}) leads to $\theta(x)=0$ and $2C_{1}+C_{0}k=0$. This last relation allows us to fix the value of the parameter $\lambda_{1}$ in equation (\ref{Sec-5-14}). By doing this we get
\begin{equation}
\lambda_{1}=\frac{\left(D^{2} -5D+5\right)\left(D^{2}-7D+13\right)}{\left(D-4\right)^{2}\left(D-2\right)\left(D-1\right)}.\label{Sec-5-15}
\end{equation}
With this, all the parameters of the theory were fixed and, fortunately, with both sectors localized. Looking at this value of  $\lambda_{1}$ we see that is not possible to localize both sectors in the case (iii) discussed in sections (\ref{Sec-3-2}) and (\ref{Sec-3-4}). The reason is that in case (iii) we have $\lambda_{1}=0$ and this is possible in Eq. (\ref{Sec-5-15}) only for non-integer values of $D$.

\begin{figure}[ht]
\centering
\includegraphics[width=.47\linewidth]{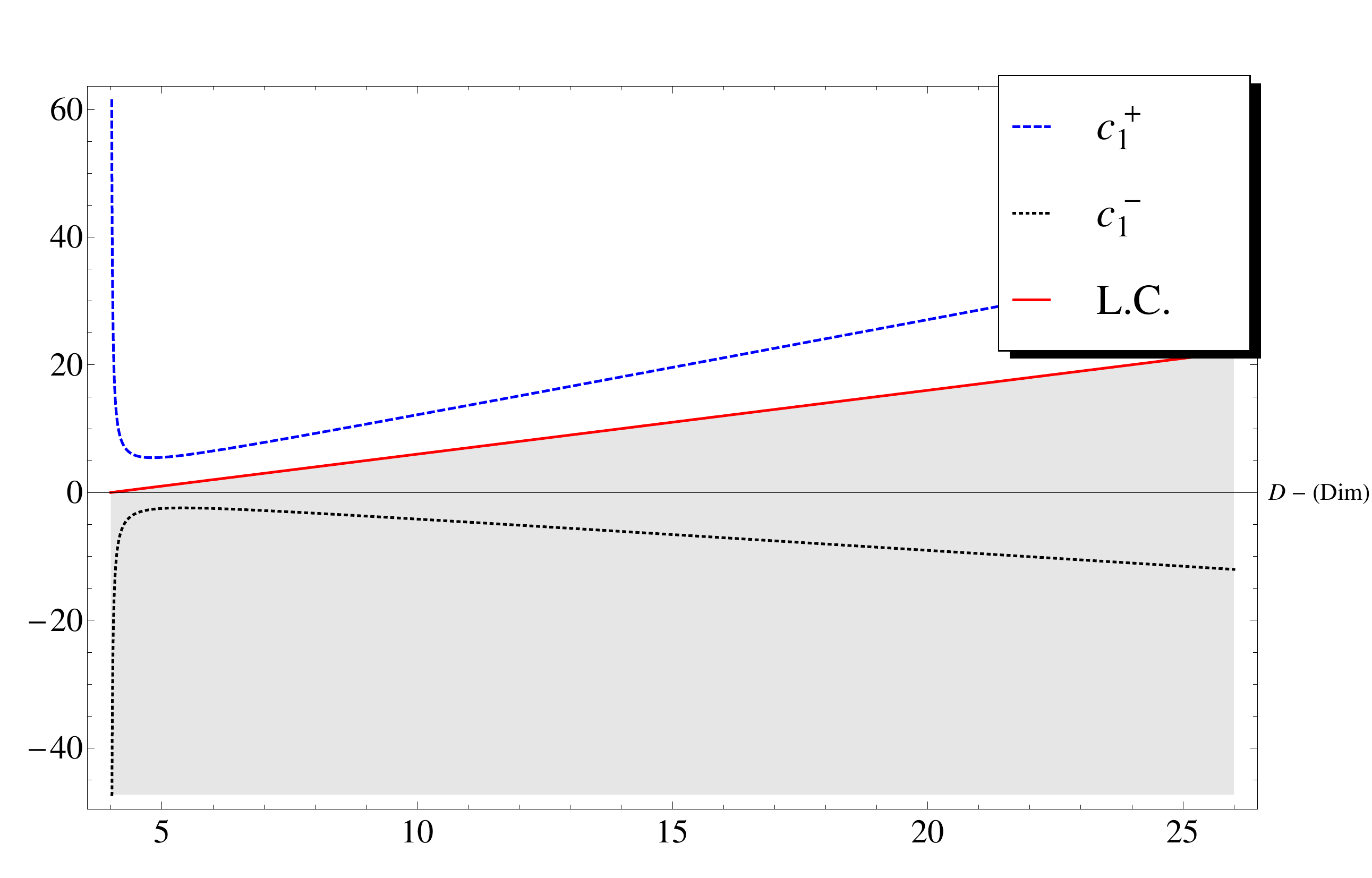}
\caption{Comparative plot of solutions  $c_{1}^{(+)}$ (dashed blue line), $c_{1}^{(-)}$ (doted black line) and the localization condition - L.C. (red line).}\label{figura3a}
\end{figure}

\begin{figure}[ht]
\centering
\includegraphics[width=.47\linewidth]{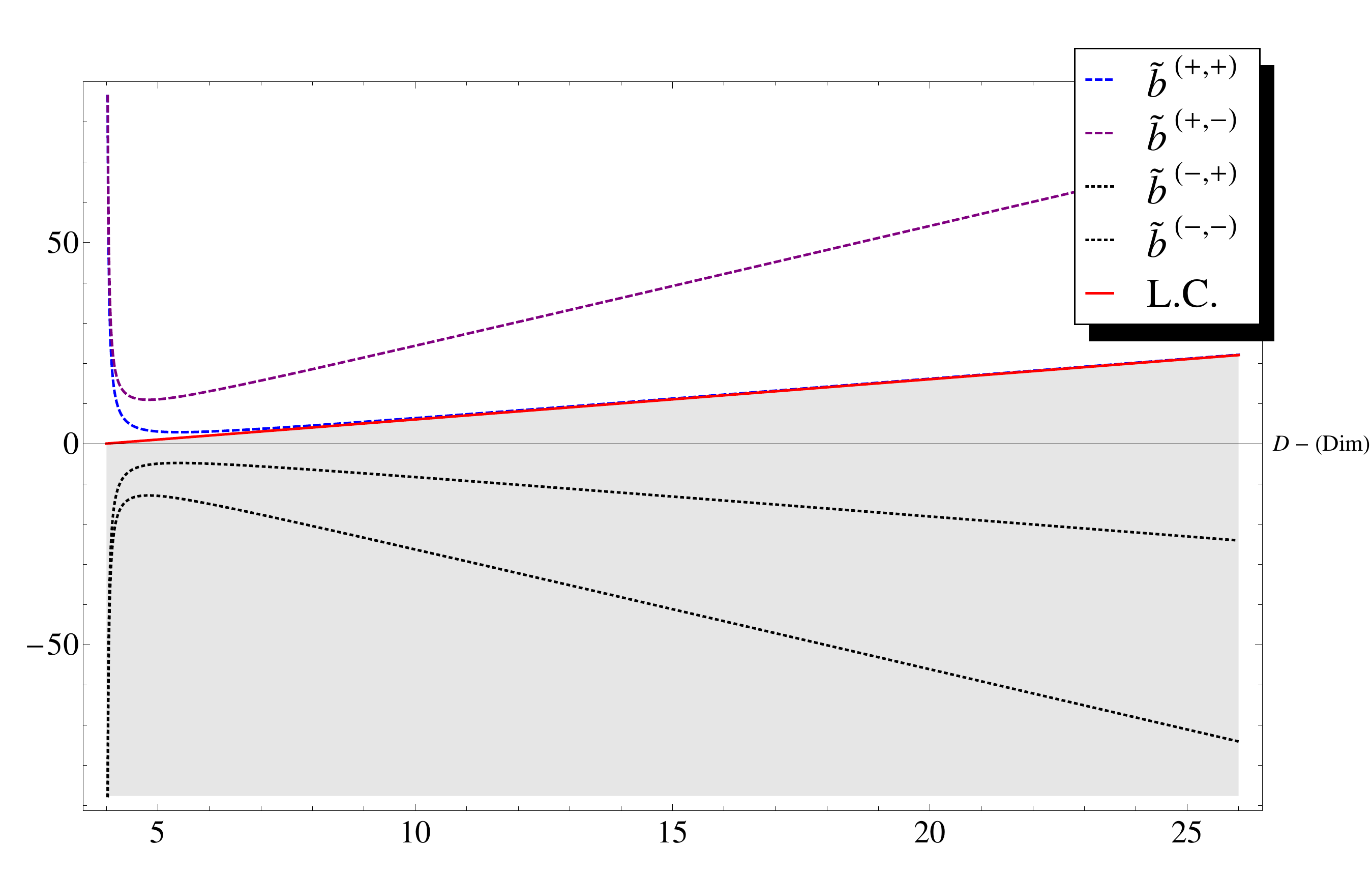}
\caption{Comparative plot of solutions  $\tilde{b}^{(+,+)}$ (dashed blue line), $\tilde{b}^{(+,-)}$ (dashed purple line), $\tilde{b}^{(-,\pm)}$ (dotted black line) and the localization condition - L.C. (red line).}\label{figura3b}
\end{figure}

Now, the localization of the gauge and the scalar fields for the case (iv) of sections (\ref{Sec-3-2}) and (\ref{Sec-3-4}) can be performed. Since both solutions are power of the warp factor as follows $e^{a\sigma(y)}$. We will carry out the analysis of the values of $a$ obtained previously for both sectors. 
\begin{itemize}
\item[(1)] {\it Gauge field} - Case (iv) of section (\ref{Sec-3-2}): for this case, the value of $\lambda_{1}$ showed in Eq. (\ref{Sec-5-15}) gives us
 \begin{equation}
  a=c_{1}^{(\pm)}=\frac{D-2}{2}\pm\sqrt{\lambda_ {1}(D-1)(D-2)+1}=\frac{D-2}{2}\pm\frac{(D-3)^{2}}{(D-4)}.\label{Sec-5-16}
 \end{equation}
 Thus, as discussed in section (\ref{Sec-3-2}), we have two solutions for the gauge field. In Fig. (\ref{figura3a}), we have a plot (solid red line) of the localization condition $2c_{1}>d=D-4$. In addition, we also have the two solutions (\ref{Sec-5-16}):  $c_{1}^{(+)}$ (dashed blue line) and $c_{1}^{(-)}$ (dotted black line). From this graphic, it is clear that the only solution that satisfies the localization condition is $c_{1}^{(+)}$. This conclusion agree with the general results obtained in (\ref{Sec-3-2}).
 \item[(2)] {\it Scalar fields} - Case (iv) of section (\ref{Sec-3-4}): for this case, the value of $\lambda_{1}$ given by Eq. (\ref{Sec-5-15}) allows us to write
 \begin{equation}
  a=\tilde{b}^{(\pm,\pm)}\!=\!-\frac{1}{2}\pm\frac{1}{2}\!\left[(D-3)^{2}\!-\!4(D-1)(\lambda_{2}^{(\pm)}+\lambda_{1}D)\right]^{\frac{1}{2}}.\label{Sec-5-17}
 \end{equation}
 Where $\tilde{b}^{(\pm,+)}=-\frac{1}{2}\pm\sqrt{...\lambda_{2}^{(+)}}$ and $\tilde{b}^{(\pm,-)}=-\frac{1}{2}\pm\sqrt{...\lambda_{2}^{(-)}}$. In this way, we have four solutions for each scalar field. 
 We should point that the localization condition for the scalar components found after the equation (\ref{Sec-3-3-18}) was obtained considering $\theta_{0}(x)\neq 0$. However the boundary conditions (\ref{Sec-5-08})-(\ref{Sec-5-12}) led us to deduce that $\theta_{0}(x)$ is zero, so we have to consider a change in the localization condition, instead of $2\tilde{b}> d+2$, we should consider  $2\tilde{b}>d$. It is easily verified by put the solution (\ref{Sec-3-3a-14a}), with $\theta_{0}(x)= 0$, in Eq. (\ref{Sec-3-3-18}). The Fig. (\ref{figura3b}) shows a plot (solid red line) of the above new localization condition and the four solutions (\ref{Sec-5-17}): $\tilde{b}^{(+,+)}$ (dashed blue line), $\tilde{b}^{(+,-)}$ (dashed purple line) and $\tilde{b}^{(-,\pm)}$ (dotted black lines). Again, as showed graphically, only the solutions $\tilde{b}^{(+,\pm)}$ can satisfy the localization condition. Therefore only in these cases the scalar field is localized on the $3$-brane.
\end{itemize}
It is important to stress that the localization of both sectors was only possible in this model because the interaction terms with the tensor and the Ricci scalar were present. Beyond this, a complete solution for the scalar components is model-dependent, therefore it is not possible ensure that the scalar components will be really confined for others warp factors. 

To conclude,  let us obtain the effective theory of these scalar components on the $3$-brane. 
 In the discussion of last paragraph, we obtained that $\theta(x)$ is zero, so the solution (\ref{Sec-3-3a-14a}) is given by 
 \begin{equation}
\mathcal{B}_{0}^{j}(x,y)=e^{-\frac{(D-2)}{2}\sigma(y^{j})}B^{j}(x,y)=\mbox{sgn}(y^{j})e^{\left(\tilde{b}^{(+)}-\frac{(D-4)}{2}\right)\sigma(y^{j})}\bar{B}^{j}(x).\label{Sec-5-18}
\end{equation}
With this, we obtain the following effective action
\begin{eqnarray}
S_{0}[\mathcal{B}^{j}]&=&-\frac{1}{2}\int d^{4}x d^{d}y e^{(D-4)\sigma(y^{j})}\partial_{\mu}\mathcal{B}_{0j}\partial^{\mu}\mathcal{B}^{j}_{0}\nonumber \\
&=&-\frac{1}{2}\int d^{d}y e^{2\tilde{b}^{(+)}\sigma(y^{j})}\int d^{4}x\partial_{\mu}\bar{B}^{j}(x)\partial^{\mu}\bar{B}_{j}(x)\label{Sec-5-19}
\end{eqnarray}
with the integrals over the extra dimensions finite. In addition, we must impose the constraint
\begin{eqnarray}
\sum^{d}_{j=1}\bar{B}^{j}(x)=0.\label{Sec-5-22}
\end{eqnarray}
In the particular case of $D=7$ and $d=3$, Eq. (\ref{Sec-5-19}) can be written as
\begin{eqnarray}
S_{0}\sim &-&\frac{1}{2}\int d^{4}x\left[\partial^{\mu}\bar{B}_{0}^{1}\partial_{\mu}\bar{B}_{0}^{1}\!+\!\partial^{\mu}\bar{B}_{0}^{2}\partial_{\mu}\bar{B}_{0}^{2}\!+\!\partial^{\mu}\bar{B}_{0}^{3}\partial_{\mu}\bar{B}_{0}^{3}\right].\label{Sec-5-23}
\end{eqnarray}
Now, we can use (\ref{Sec-5-22}) and redefine the fields such that the effective action for the zero mode of the scalar field if given by
\begin{eqnarray}
S_{0}\sim &-&\int d^{4}x\left[\frac{1}{2}\partial^{\mu}\bar{\bar{B}}_{0}^{1}(x)\partial_{\mu}\bar{\bar{B}}_{0}^{1}(x)\!+\! \frac{1}{2}\partial^{\mu}\bar{\bar{B}}_{0}^{2}(x)\partial_{\mu}\bar{\bar{B}}_{0}^{2}(x)\right].\label{Sec-5-24}
\end{eqnarray}
That is the action for free scalar fields.

\section{Final Remarks}\label{Sec-6}
In this work was shown that the geometric coupling has an universal validity as localization mechanism for a $D$-vector field on braneworlds with arbitrary co-dimension. It was considered a $D$ dimensional bulk with a generic conformally flat metric $e^{2\sigma}(\eta_{\mu\nu}dx^{\mu}dx^{\nu}+\delta_{ij}dy^{i}dy^{j})$, which the warp factor depends only on the transverse extra dimensions $\sigma(y)$. In this context it was proposed interaction terms of the $D$-vector field $\mathcal{A}_{N}=\left(\mathcal{A}_{\mu},\mathcal{B}_{k}\right)$ with the scalar and the Ricci tensor, and we showed that it allows us to obtain an effective theory for a gauge field and also one for scalar fields. The study of zero-mode localization for the fields $\mathcal{A}_{\mu}$ (gauge field) and $\mathcal{B}_{k}$ (scalar fields) was separated in some particular cases: (a) non-minimal coupling only with \textit{Ricci scalar}; (b) only with \textit{Ricci tensor}; and (c) the case with both interaction terms.

In section (\ref{Sec-3-1}), we analyzed the abelian vector field problem, where the features of the background geometry allowed us to obtain a Schrödinger-type equation with the potential given by (\ref{Sec-3-09a}). Such equation has a general analytic solution for the massless mode given by a function of the warp factor, $\chi_{0}^{(gauge)}=e^{a\sigma(y)}$, which is a solution when $a^{2}=c_{1}^{2}=c_{2}$, with $c_{1}$ and $c_{2}$ presented in Eq. (\ref{Sec-3-09d}). Due to its shape, this solution becomes valid for a wide variety of warp factors, either for delta-like or smooth branes. Furthermore, the mere existence of this zero-mode solution excludes any possible tachyonic mode of the theory. This is a powerful result, because, unlike other localization methods, the geometric coupling does not depend of the specific braneworld model, but only of the properties of spacetime itself. Among the main aspects, the localization condition $(2c_{1}>d)$ can be obtained by imposing only that the background is an asymptotically AdS spacetime. In section (\ref{Sec-3-2}) was made a detailed analysis of the cases (a)-(c) above mentioned, where we obtained: For case (a), the analytic solution is given by $\chi_{0}^{(gauge)}=e^{\sigma(y)}$, with $a=c_{1}=1$ and the coupling constant given by $\lambda_{1}=\frac{D-6}{4 (D-1)}$. Therefore, from the localization condition, such zero-mode solution is confined when $d<2$, \textit{i.e.}, in co-dimension one models. For case (b), we get the analytic solution $\chi_{0}^{(gauge)}=e^{\frac{D}{2}\sigma(y)}$ and the coupling constant $\lambda_{2}=-2$. This solution is localized for any number of transverse extra dimensions, since the localization condition becomes $d<D$ and can always be satisfied. Finally for case (c), the analytic solution is obtained when $a=c^{(\pm)}_{1}=\frac{D-2}{2}\pm\sqrt{\lambda_ {1}(D-1)(D-2)+1}$, with $\lambda_ {1}$ a 'free' parameter. This solution is also localized on a $3$-brane for any number of extra dimensions, provided that the parameter $\lambda_ {1}$ belongs to the range $(-(D-1)^{-1}(D-2)^{-1},0)$ for $c^{(-)}_{1}$, or to the range $(-(D-1)^{-1}(D-2)^{-1},\infty)$ for $c^{(+)}_{1}$. To conclude this discussion about the effective gauge field, we obtained some conditions for which the localization of gravity ensures the localization of gauge field sector in this scenarios. As discussed at the and of the section (\ref{Sec-3-2}), the zero-mode solution obtained in Ref. \cite{Csaba} for the gravitational field is given by $\psi_{0}^{(gravity)}=e^{\frac{(D-2)}{2}\sigma}$. When we compare this solution to the above analysis for the gauge field, we can see that if the coupling of vector field with the Ricci tensor is present in the theory, then the localization of gravity is enough to ensure the localization of gauge field. This is an interesting result, since that all the consistency of the model will depend only on the fact that gravity is consistent.

We also considered the localization of the scalar components $\mathcal{B}^{j}$ of the $D$-vector field $\mathcal{A}_{M}$. As said in the introduction, in co-dimension one models the scalar component is never localized simultaneously with the gauge field sector. This is a drawback since the backreaction of this field could alter the AdS vacuum. In section (\ref{Sec-3-3}), we studied this problem and showed  that when more co-dimensions are considered there is an indication that such scalar components can be localized simultaneously with the gauge field sector $A_{\mu}$. Differently of this sector, a general analytical treatment of the $\mathcal{B}^{j}$ was not found. This is due to the fact that the equation of motion (\ref{Sec-3-3-03}) can not be diagonalized and therefore are always coupled. However, as we are interested in convergence conditions of the solution, an asymptotic treatment was performed for the cases (a)-(c) above. With this in mind the asymptotic solutions was found in Eq. (\ref{Sec-3-3a-14a}). This solution indicates that the localization of the gauge and scalar components of the vector field can be simultaneously obtained only for the cases (b) and (c). Therefore only when the interaction with the Ricci tensor is switched on, as showed in fig. (\ref{figura2a}) and (\ref{figura2b}). We should stress that this is another important result of this work, since the localization of both components ensures that the backreaction of $\mathcal{B}^{j}$ will not jeopardise the AdS feature of the vacuum. Moreover, the localization of these components does not imply any modification on the Coulomb law, instead they could be interpreted as Higgs fields, or even dark energy. However, we could not ensure that these components are really confined because there is no guarantee that the solutions will be regular in all range of integration. In order to fully solve the scalar components problem, a specific background must be considered.

As an application, the intersecting brane model, developed by Arkani-Hamed {\it et al}, was used to discuss in a more detailed way these general results presented by us. Beyond that, it was possible to obtain the complete solutions for scalar components given by (\ref{Sec-5-18}). As said before, the branes introduce a set of boundary conditions that fix all the parameters of the model. With this, we verified that the localization of the scalar and gauge components is not possible for the case (b) above. Therefore a fully consistent model is possible only for the case (c), when couplings with the \textit{scalar} and the \textit{Ricci tensor} are considered. Furthermore, in Eq. (\ref{Sec-5-24}) we found the effective field theory for the scalar sector at the $3$-brane, which are free massless scalar fields. With this we get that the our final effective action has one free gauge field plus $(d-1)$ free scalar fields. The scalar fields can play an important role in cosmology and particle physics and in principle can provides phenomenological consequences of the geometrical localization mechanism. We should also point that we have not considered backreactions of fluctuations of the geometry. To study this is very important in order to understand if these effects will destroy the universality found at the first level. However, in the moment, these aspects are beyond the scope of this paper and can be treated in a future work. 

\section*{Acknowledgement}
The authors would like to thanks Alexandra Elbakyan and sci-hub, for removing all barriers in the way of science.
We acknowledge the financial support provided by the Conselho Nacional
de Desenvolvimento Científico e Tecnológico (CNPq) and Fundação Cearense de Apoio
ao Desenvolvimento Científico e Tecnológico (FUNCAP) through PRONEM PNE-0112-
00085.01.00/16.

\appendix
\section{Appendix}\label{Apendice}
\subsection{Splitting of the Action (\ref{Sec-3-02}) in the Sectors $S_{\perp}[\hat{\mathcal{A}}_{\mu}]$ and $S[\phi,B_{k}]$}\label{Apendice2}
In section (\ref{Sec-3-1}), we proposed the separation $\mathcal{A}_{N}=\left(\hat{\mathcal{A}}_{\mu}+\partial_{\mu}\phi,B_{k}\right)$, where $\partial_{\mu}\hat{\mathcal{A}}^{\mu}=0$ is the transverse sector of the effective abelian vector field on the brane. This proposal allowed carry out the separation of action (\ref{Sec-3-02}) in one part containing only the transverse sector and another part with longitudinal and scalar components $B_{k}$ sectors. Here, we will only clarify this procedure.

From the action (\ref{Sec-3-02}) we have the kinetic term in $D$-dimensions, 
\begin{eqnarray}
\frac{1}{4}\mathcal{F}_{MN}\mathcal{F}^{MN}=\frac{1}{4}\hat{\mathcal{F}}_{\mu\nu}\hat{\mathcal{F}}^{\mu\nu} +\frac{1}{4}\mathcal{B}_{jk}\mathcal{B}^{jk}+\frac{1}{2}\mathcal{F}_{\mu j}\mathcal{F}^{\mu j},\label{A-04}
\end{eqnarray}
where $(\mu, \nu,...)$ are related to the brane coordinates and $(j,k,...)$ are related to extra dimensions. The first term in expression (\ref{A-04}) is already written as a function only of the transverse sector $\hat{\mathcal{A}}^{\mu}$. The last term, which still has coupled terms of transverse sector with the other sectors, can be written as follows
\begin{eqnarray}
\frac{1}{2}\mathcal{F}_{\mu j}\mathcal{F}^{\mu j}&=&\frac{1}{2}g^{\mu\nu}g^{jk}\left(\partial_{\mu}\mathcal{B}_{j}-\partial_{j}\hat{\mathcal{A}}_{\mu}-\partial_{j}\partial_{\mu}\phi\right)\left(\partial_{\nu}\mathcal{B}_{k}-\partial_{k}\hat{\mathcal{A}}_{\nu}-\partial_{k}\partial_{\nu}\phi\right)\nonumber\\
&=&\frac{1}{2}g^{\mu\nu}g^{jk}\partial_{\mu}\mathcal{B}_{j}\partial_{\nu}\mathcal{B}_{k}-g^{\mu\nu}g^{jk}\partial_{\mu}\mathcal{B}_{j}\partial_{k}\partial_{\nu}\phi+\frac{1}{2}g^{\mu\nu}g^{jk}\partial_{j}\partial_{\mu}\phi\partial_{k}\partial_{\nu}\phi\nonumber\\
&+&\frac{1}{2}g^{\mu\nu}g^{jk}\partial_{j}\hat{\mathcal{A}}_{\mu}\partial_{k}\hat{\mathcal{A}}_{\nu}+g^{\mu\nu}g^{jk}\partial_{j}\hat{\mathcal{A}}_{\mu}\partial_{k}\partial_{\nu}\phi-g^{\mu\nu}g^{jk}\partial_{\mu}\mathcal{B}_{j}\partial_{k}\hat{\mathcal{A}}_{\nu}.\label{A-05}
\end{eqnarray}
The last two terms in this relation can be converted in boundary terms (in the coordinates of the brane) due to the condition $\eta^{\mu\nu}\partial_{\nu}\hat{\mathcal{A}}_{\mu}=0$. Thus, if we consider that the boundary terms are zero, then the action (\ref{Sec-3-02}) can be written as follows
\begin{equation}
 S=S_{\perp}[\hat{\mathcal{A}}_{\mu}]+S\left[\phi,\mathcal{B}_{k}\right],\label{A-06}
\end{equation}
where
\begin{eqnarray}
\! \! S_{\perp}=-\!\!\int \! d^{(D-d)}x d^{d}y \sqrt{-g}\left\lbrace\frac{1}{4}g^{\mu\nu}g^{\rho \lambda}\hat{\mathcal{F}}_{\mu\rho}\hat{\mathcal{F}}_{\nu\lambda}+\frac{1}{2}g^{\mu\nu}g^{jk}\partial_{j}\hat{\mathcal{A}}_{\mu}\partial_{k}\hat{\mathcal{A}}_{\nu}\right.\nonumber\\+\left.\frac{\lambda_{1}}{2}Rg^{\mu \nu}\hat{\mathcal{A}}_{\mu}\hat{\mathcal{A}}_{\nu}+\frac{\lambda_{2}}{2}g^{\mu\nu}g^{\rho \lambda}R_{\mu\rho}\hat{\mathcal{A}}_{\nu}\hat{\mathcal{A}}_{\lambda}\right\rbrace\label{A-07}
\end{eqnarray}
and
\begin{eqnarray}
\! \! S\left[\phi,\mathcal{B}_{k}\right]&=&-\!\!\int \! d^{(D-d)}x d^{d}y \sqrt{-g}\left\lbrace\frac{1}{4}g^{jm}g^{kl}\mathcal{B}_{jk}\mathcal{B}_{ml}+\frac{1}{2}g^{\mu\nu}g^{jk}\partial_{\mu}\mathcal{B}_{j}\partial_{\nu}\mathcal{B}_{k}\right.\nonumber\\&-&g^{\mu\nu}g^{jk}\partial_{\mu}\mathcal{B}_{j}\partial_{k}\partial_{\nu}\phi+\frac{1}{2}g^{\mu\nu}g^{jk}\partial_{j}\partial_{\mu}\phi\partial_{k}\partial_{\nu}\phi \nonumber\\&+&\left.\!\frac{\lambda_{1}}{2}Rg^{jk}\mathcal{B}_{k}\mathcal{B}_{j}\!+\!\frac{\lambda_{2}}{2}g^{lm}g^{jk}R_{jl}\mathcal{B}_{m}\mathcal{B}_{k}\right\rbrace.\label{A-08}
\end{eqnarray}

\end{document}